\documentclass[useAMS,usenatbib]{mn2e}
\usepackage{graphicx}
\usepackage{lscape}

\title[Optical Fe\,{\sc iii} lines]{Iron abundances from optical Fe\,{\sc iii} absorption lines in B-type stellar spectra}

\author[H. M. A. Thompson et al.]
	 {H. M. A. Thompson$^{1}$\thanks{email: h.thompson@qub.ac.uk
	\hspace*{0.3cm}
	 \newline Based on observations made with the WHT operated on the island of La Palma by the Isaac Newton Group in the Spanish Observatorio del Roque de los Muchachos of the Instituto de Astrof\'{i}sica de Canarias},
	 F. P. Keenan$^{1}$, P. L. Dufton$^{1}$, C. Trundle$^{1}$, R. S. I. Ryans$^{1}$ 
\newauthor and P. A. Crowther$^{2}$
	\\
       	$^{1}$Astrophysics Research Centre, School of Mathematics and Physics, Queen's University, Belfast 
BT7 1NN\\
	$^{2}$Department of Physics and Astronomy, University of Sheffield, Hicks Building, Hounsfield Rd, Sheffield, S3 7RH\\
       }

\date{Accepted  
       Received 
       in original form }

\pagerange{\pageref{firstpage}--\pageref{lastpage}} \pubyear{0000}
 
\def\LaTeX{L\kern-.36em\raise.3ex\hbox{a}\kern-.15em
    T\kern-.1667em\lower.7ex\hbox{E}\kern-.125emX}

\begin{document}

\label{firstpage}

\maketitle

\begin{abstract}
The role of optical Fe\,{\sc iii} absorption lines in B-type stars as iron abundance diagnostics is considered. To date, ultraviolet Fe lines have been widely used in B-type stars, although line blending can severely hinder their diagnostic power. 
Using optical spectra, covering a wavelength range $\sim$ 3560 -- 9200 \AA, a sample of Galactic B-type main-sequence and supergiant stars of spectral types B0.5 to B7 are investigated. 
A comparison of the observed Fe\,{\sc iii} spectra of supergiants, and those predicted from the model atmosphere codes {\sc tlusty} (plane-parallel, non-LTE), with spectra generated using {\sc synspec} (LTE), and {\sc cmfgen} (spherical, non-LTE), reveal that non-LTE effects appear small. 
In addition, a sample of main-sequence and supergiant objects, observed with FEROS, reveal LTE abundance estimates consistent with the Galactic environment and previous optical studies. 
Based on the present study, we list a number of Fe\,{\sc iii} transitions which we recommend for estimating the iron abundance from early B-type stellar spectra.
\end{abstract}


\section{Introduction}

Iron lines dominate the spectra of many astrophysical objects, such as novae \citep{{mck97},{hat07}}, photoionized H II regions \citep{{rub97},{rod02},{est02}} and active galactic nuclei \citep*{{sig03},{sig04},{zha06}}. 
The atomic processes for iron and other iron-group ions have been the subject of numerous investigations, for example the IRON Project \citep{hum93} which considers applications in astrophysical and laboratory plasmas \citep{pra96a}. 
Absorption lines of iron provide important metallicity diagnostics for both stars and galaxies, and also play a key role in investigating star formation histories through element ratios, such as [$\alpha$/Fe] \citep{gil98}. However, there is a lack of robust Fe abundance determinations in external galaxies,  e.g. the Magellanic Clouds (see for example, \citealt*{{rol02},{tru02},{tru07}} and \citealt{mok07}), due to the complex Grotian diagrams for Fe, the low metallicity environment of the Magellanic Clouds and the reliability of the currently available atomic data. 

Early B-type stars are important for studying the chemical composition of our own and other galaxies \citep{{kil92},{duf98}}. 
In the optical spectra of B-type stars, iron lines due to a number of ionization stages are observed (see for example, \citealt*{{gie92},{len93},{sma97},{mor06}}). Absorption features arising from Fe\,{\sc iii} are primarily detected \citep{har70}, with Fe\,{\sc ii} also found in later B-type stars \citep{pin93}. Fe\,{\sc iv} lines are not expected in the optical spectra of B-type stars due to their intrinsic weakness in this temperature range.

The optical Fe\,{\sc iii} line spectrum has not been widely employed to determine abundances, as it has often been believed to be too weak to provide reliable measurements \citep{ken94a}. 
However, it has been used for chemical composition studies of several bright, narrow-lined main-sequence B-type stars, such as $\zeta$\,Cas, $\gamma$\,Peg, $\iota$\,Her, $\tau$\,Sco and $\lambda$\,Lep \citep{{sni69},{pet70},{har70},{pet76},{pet85}}.
These Galactic objects have sufficiently high Fe content, coupled with narrow metal absorption lines (due to their low projected rotational velocities), so that, even with relatively poor quality optical spectra, Fe\,{\sc iii} features can be detected. 
However, the quality of the available spectra did not allow the study of Fe\,{\sc iii} lines in objects of lower metallicity. 
Therefore, more recently the optical wavelength range has been largely overlooked in favour of the ultraviolet domain \citep{{swi76},{pet85},{dix98}}, and in particular the very rich spectral region around 1900 \AA\ \citep*{{tho74},{heb83},{ken94a},{gri96},{moe98}}.  
On the other hand, due to the high density of absorption features and resultant blending, continuum placement in the ultraviolet is difficult, and hence significant errors may be present in the derived abundances \citep{moe98}. 

Due to instrumental advances in more recent years, it has become routine to obtain high resolution and signal-to-noise (S/N) spectra, and the use of the optical Fe\,{\sc iii} lines as a diagnostic has been revisited (e.g. \citealt{{cun94},{kil94}}). 
A number of main-sequence objects have subsequently been re-examined, for example $\gamma$\,Peg and $\iota$\,Her \citep*{{pin93},{zon98}}, producing results consistent with the earlier optical analyses. 
In addition, the higher S/N ratios of the available spectra have made it possible to detect the weak optical Fe\,{\sc iii} lines in a number of lower metallicity objects, such as AV\,304 in the Small Magellanic Cloud \citep{rol03}, and the globular cluster post-AGB stars ZNG-1 in M\,10 \citep{moo04}, Barnard\,29 in M\,13 and ROA\,5701 in $\omega$\,Cen \citep{tho07}.

A number of B-type stars have been studied using both ultraviolet and optical spectra, but the corresponding abundance estimates are generally in poor agreement. Estimates from the ultraviolet spectra are consistently lower than those from the optical, for example in the post-AGB stars Barnard\,29, ROA\,5701 \citep{tho07}, BD\,+33${^\circ}$2642 \citep*{nap94} and HD\,177566 \citep{{ken94a},{ken94b}}. Young Galactic objects  including $\gamma$\,Peg \citep{moe98} and $\iota$\,Her \citep{gri96}, plus stars in known metallicity environments such as the Magellanic Clouds \citep{duf07}, also display similar discrepancies.

Here, high resolution spectra for a number of narrow-lined Galactic B-type main-sequence and supergiant stars, covering a range of spectral types, are analysed using LTE and non-LTE model atmosphere techniques. Details on the observations, models and Fe\,{\sc iii} line selection can be found in Sections \ref{sec_obs} and \ref{sec_data}.  
An LTE approximation is considered, and the reliability of such an assumption, along with an assessment of the individual Fe\,{\sc iii} lines, is discussed in Section \ref{sec_discuss}.

\section{Observations and data reduction}
\label{sec_obs}

The observational data included here are used for two investigations. Firstly, to allow a comparison between the {\sc tlusty} \citep*{{hub88},{hub95},{hub98}} and {\sc cmfgen} \citep{hil03} model atmosphere codes, and secondly, an investigation of the reliability of the Fe\,{\sc iii} spectrum to estimate abundances. The first investigation utilizes a subset of the stars discussed in \citet*{cro06} for which there were suitable data from ISIS/WHT and FEROS/ESO, and these objects are listed in Table \ref{tab_tca}. \citet{cro06} analysed a sample of 25 Galactic B-type supergiants using optical spectra from three sources; CTIO data obtained using the Loral\,1\,K spectrograph, WHT spectra taken using the ISIS spectrograph (Smartt, private communication) and JKT spectra from the Richardson-Brealey spectrograph \citep*{len92}. 
A number of objects were omitted from this study due to having poorly detected or unseen Fe\,{\sc iii} features (i.e. objects of spectral type B0.5 and earlier), or if their atmospheric parameters lay outside the available {\sc tlusty} grid points (for example, HD\,13854, HD\,152236, HD\,190603 and HD\,194279).
For the second investigation, a sample of B-type stars, consisting of two main-sequence stars and nine supergiants, has been selected, from a FEROS/ESO dataset. These are analysed using the {\sc tlusty} atmosphere code and are listed in Table \ref{tab_atparam}. The objects have been chosen to cover a range of spectral types and have relatively narrow-lined absorption spectra. 
The supergiant HD\,53138 has been included in both samples due to it having a rich Fe\,{\sc iii} line spectrum.

The WHT data were taken using the Intermediate-dispersion Spectrograph and Imaging System (ISIS), on 1999 July 08. The EEV12 CCD detector (2148$\times$4200 pixels) was used on the blue arm with the R1200 grating and a slit width of 1.0 arcsec. This resulted in a wavelength coverage of $\sim$ 3920--4720 \AA. Typical S/N values for spectra at $\sim$ 4500 \AA\ are given in Table \ref{tab_tca}. 
The data were reduced using standard procedures within {\sc iraf}, which included flat-fielding, bias subtraction and wavelength calibration and the combined spectra read into the Starlink package {\sc dipso} \citep{how04} for further analysis.

High resolution ESO spectra were obtained using the Fiber fed Extended Range Optical Spectrograph (FEROS, \citealt{kau99}), between 2005 April 21 and 24 (proposal 075.D-0103(A)). The EV 2K$\times$4K CCD detector and 79 lines mm$^{-1}$ \'{e}chelle grating were used, resulting in a spectral resolution of $R$ $\sim$ 48000 and a wavelength coverage of $\sim$ 3560--9200 \AA.   
Data were reduced using the standard FEROS Data Reduction System (DRS), implemented with ESO--{\sc midas} software. The reduced spectra for a given target were subsequently combined using {\sc scombine} within {\sc iraf}\footnote{{\sc iraf} is distributed by the National Optical Astronomy Observatories, which are operated by the Association of Universities for Research in Astronomy, Inc., under cooperative agreement with the National Science Foundation.}, and imported into {\sc dipso} for further analysis. Their average S/N ratios at $\sim$ 4500 \AA\ are given in Tables \ref{tab_tca} and \ref{tab_atparam}.
  
The spectra were normalised with low-order polynomials and radial velocity shifted to the rest frame using unblended metal and hydrogen lines. Absorption line equivalent widths were estimated by fitting Gaussian profiles to the lines. The error in measuring these is discussed in \citet{duf90}, with a well-observed unblended feature accurate to 10 per cent (designated `a' in Tables \ref{tab_tcab}, \ref{tab_sab} and \ref{tab_mab}), weak or blended features accurate to 20 per cent (designated `b') and weaker features accurate to less than 20 per cent (designated `c').

\section{data analysis}
\label{sec_data}
\subsection{Model atmosphere techniques}
\label{sec_model}

Two model atmosphere codes have been used in this paper, namely {\sc tlusty} and {\sc cmfgen}, which are discussed below.

Non-LTE model atmosphere grids, calculated with {\sc tlusty} and {\sc synspec} \citep{{hub88},{hub95},{hub98}}, have been used to derive atmospheric parameters. The methods involved are discussed in, for example, \citet{rya03}, \citet{hun05}, \citet{{tho06},{tho07}} and \citet{tru07}, whilst greater detail about the grids may be found in \citet{duf05} and at http://star.pst.qub.ac.uk.
Briefly, four grids were calculated corresponding to the Galaxy, the Large (LMC) and Small (SMC) Magellanic Clouds and a lower metallicity regime. The Galactic grid has been adopted throughout this analysis and covered a range of effective temperatures (12000 to 35000 K), logarithmic gravities (log~{\em g} = 4.5 dex down to the Eddington limit; {\em g} in cm\,s$^{-2}$) and microturbulences ($\xi$ = 0, 5, 10, 20 and 30 ${\rm km\,s^{-1}}$). Typical spacing between grid points for the effective temperature and surface gravities were 2500 K and 0.25 dex, respectively. 
For accurate profile fitting of the hydrogen lines, further models were executed at a resolution of 0.1 dex for log~{\em g}, with these models used only in determining the surface gravity. 
The metal line blanketing was assumed to be dominated by iron, with the exclusion of nickel in our models unlikely to be a source of significant error in the atmospheric parameters and chemical composition \citep{hub98}.

Iron is included in the non-LTE {\sc tlusty} model atmosphere calculations with sets of ionic levels being collected into superlevels in order to make the calculations tractable. Although in the spectral synthesis code it is possible to assign ionic levels to appropriate superlevels, we have instead opted to calculate the Fe\,{\sc iii} spectrum in an LTE approximation. 
These were then employed with the measured equivalent widths to derive the abundance estimates shown in Tables \ref{tab_tcab}, \ref{tab_sab} and \ref{tab_mab} (see \citealt{tho07}). 
Atomic data for the Fe\,{\sc iii} lines were taken from \citet{nah96}, as these are (to our knowledge) the most up-to-date values available. 
Their oscillator strengths are shown in Table \ref{tab_atdata} (available in full on-line), along with those from the Kurucz database (http://nova.astro.umd.edu, \citealt{{hub88},{hub95},{hub98}}), plus the wavelength, transition, and g$_{i}$ and g$_{j}$ values for the selected iron lines.

The {\sc tlusty} model atmosphere grids are based upon the standard solar composition \citep{gre98}, while {\sc synspec} allowed a range of iron abundances to be calculated. Thus, using the Galactic grid, spectra were synthesised for a range of iron abundances from 6.7 and 8.3 dex. 
\citet{duf07} discuss the possible inconsistencies which can arise from the adopted abundances used in both the model atmosphere and spectrum synthesis calculations, i.e. the use of {\sc synplot} to synthesize spectra for a range of metallicities which differ from that of the underlying model atmosphere calculation. 
Tests discussed by \citet{duf07} indicate that these should not be a significant source of error, and the method has been widely used (see for example \citealt*{{kil94},{mce99}} and \citealt{kor00}).

{\sc cmfgen} is a non-LTE line blanketed multi-purpose atmospheric code designed for the spectral analysis of stars with stellar winds. It solves the radiative transfer equation for spherical geometry in the co-moving frame, under the constraints of statistical and radiative equilibrium \citep{hil98}. As {\sc cmfgen} does not solve the momentum equation, the mass-loss rate and velocity law must be specified. 

\citet{cro06} used the current version of {\sc cmfgen} \citep{hil03} and again adopted the atomic data of \citet{nah96} for the Fe\,{\sc iii} lines. A model similar to that applied by \citet{eva04} was used by \citet{cro06}, comprising 26 ions of H, He, C, N, O, Mg, Al, Si, S, Ca and Fe for early B subtypes, containing a total of 2636 individual levels which were grouped into 766 superlevels, with a full array of 25960 bound-bound transitions. Higher ionization stages were excluded for spectral types B2.5 and B3. The temperature structure was determined by radiative equilibrium, with the velocity parameterized with a classical $\beta$-type law for the supersonic part (with the exponent in the range $\beta$ = 1 -- 3, selected on the basis of the H$\alpha$ profile), and the subsonic velocity set by the corresponding H-He {\sc tlusty} photospheric models calculated for each {\sc cmfgen} model atmosphere calculation \citep{hil01}. 
Test calculations have been performed for a H-He {\sc tlusty} photospheric model and a blanketed {\sc tlusty} model (Dr. S. Searle, private communication). The results showed comparable stellar parameters, and hence the use of the H-He {\sc tlusty} model here appears valid for this analysis. 
Synthetic spectra for each star were obtained for iron abundances of Fe = 7.35, 7.65 and 7.95 dex, adopting the appropriate atmospheric parameters. These were used to derive the abundance estimates given in Table \ref{tab_tcab} (see Section \ref{sec_femethod} for further details).

\subsection{Atmospheric parameters}
\label{sec_at}

\begin{table*}
\caption{Adopted atmospheric parameters for the sample of supergiants employed in the comparison between the model atmosphere calculations using {\sc cmfgen} and {\sc tlusty}. The spectral types and {\sc cmfgen} parameters are taken from \citet{cro06}, with additional spectral types from \citet{len92}$^a$. Note that the atmospheric parameters derived from the two codes have, in general, been determined using different data sets (D$^{\dagger}$).} 
\label{tab_tca}
\begin{tabular}{@{}llccccccc@{\extracolsep{0.11cm}}ccccc}
\hline
		&		& 		&\multicolumn{6}{c}{\sc tlusty}		& \multicolumn{5}{c}{\sc cmfgen}	\\
						\cline{4-9} 				\cline{10-14} \\[0.3mm]
HD 		& Name		& Spectral	&$T_{\rm eff}$    &log~{\em g}	&$\xi$			&$\upsilon$\,sin{\it i}	&D$^{\dagger}$	&S/N 	&$T_{\rm eff}$    &log~{\em g}	&$\xi$ 		      &$\upsilon$\,sin{\it i}	&D$^{\dagger}$	\\
Number		& 		& Type		&(${\rm 10^{3}K}$)&(dex)  	&(${\rm km\, s^{-1}}$)	&(${\rm km\, s^{-1}}$) 	&		&ratio	&(${\rm 10^{3}K}$)&(dex)  	&(${\rm km\, s^{-1}}$)&(${\rm km\, s^{-1}}$) 	\\
\hline
152235		& V900\,Sco	& B0.7Ia	& 22.7 & 2.70 & 15 & 76 & 1 & 400   & 23.0 & 2.65 & 10 & 81 & 3 \\   	
154090		& V1073\,Sco	& B0.7Ia	& 22.3 & 2.70 & 16 & 80 & 1 & 420   & 22.5 & 2.65 & 10 & 78 & 3 \\   	
148688		& V1058\,Sco	& B1Ia		& 20.6 & 2.50 & 16 & 71 & 1 & 435   & 22.0 & 2.60 & 15 & 72 & 3 \\   	
 14956		& V475\,Per	& B1.5Ia$^a$	& 19.4 & 2.40 & 20 & 75 & 2 & 370   & 21.0 & 2.50 & 10 & 80 & 2 \\   	
 14818		& V554\,Per	& B2Ia$^a$	& 18.8 & 2.40 & 18 & 76 & 2 & 240   & 18.5 & 2.40 & 20 & 82 & 2 \\   	
 14143		& 		& B2Ia$^a$	& 19.1 & 2.40 & 18 & 72 & 2 & 230   & 18.0 & 2.25 & 20 & 76 & 2 \\   	
 53138		& $o^{2}$\,CMa	& B3Ia$^a$	& 15.2 & 2.00 & 18 & 48 & 1 & 330   & 15.5 & 2.05 & 20 & 58 & 4 \\   	
\hline
\end{tabular}
D$^{\dagger}$=Data source: 1 = FEROS, MPG/ESO 2.2-m telescope; 2 = ISIS, WHT; 3 = Loral\,1\,K spectrograph, CTIO; 4 = Richardson-Brealey spectrograph, JKT.  
\end{table*}

\begin{table*}
\begin{center}
\caption{Adopted atmospheric parameters for the sample stars (nine supergiants and two main-sequence stars). Spectral types are from the Bright Star Catalogue \citep{hof95}, with additional data from \citet{per97}$^a$, \citet{sch95}$^b$ and \citet{len92}$^c$. All of the objects have been observed using FEROS on the MPG/ESO 2.2-m telescope.}
\label{tab_atparam}
\begin{tabular}{@{}llccccccccccc}
\hline
HD 		& Name		& Spectral	&$T_{\rm eff}$	   & log~{\em g}	& $\xi$		  	&$\upsilon$\,sin{\it i}	&  S/N      \\
Number		& 		& Type		&(${\rm 10^{3}K}$) & (dex)	  	&(${\rm km\,s^{-1}}$)  &(${\rm km\,s^{-1}}$)	&  ratio     \\
\hline
155985		&      		& B0.5Iab$^a$	& 23.4  & 3.00 & 13   & 63   & 370	\\
108002	  	& 	 	& B1Iab$^a$ 	& 20.4  & 2.70 & 16   & 65   & 385	 \\
142758	  	& V361\,Nor 	& B1.5Ia$^a$	& 19.0  & 2.50 & 22   & 65   & 415	 \\  
141318	  	& V360\,Nor 	& B2II      	& 20.6  & 3.30 & 13   & 43   & 475       \\
 53138		& $o^{2}$\,CMa	& B3Ia$^c$	& 15.2  & 2.00 & 18   & 48   & 330   	 \\
 51309	  	& $\iota$\,CMa 	& B3II      	& 16.6  & 2.30 & 19   & 49   & 370       \\
159110		& 	 	& B4Ib$^b$	& 20.1  & 3.30 & 10   & 18   & 500	 \\
164353		& 67\,Oph 	& B5II$^c$	& 15.8  & 2.60 & 23   &  9   & 500	 \\
 91619		& V369\,Car 	& B7Iae		& 13.2  & 1.80 & 20   & 29   & 500	 \\
 \hline
126341		&$\tau^{1}$\,Lup& B2IV		& 21.5  & 3.65 & 3    & 18   & 500	 \\
 79447		& $\iota$\,Car 	& B3III		& 18.0  & 3.60 & 9    &  0   & 420	 \\	 
\hline
\end{tabular}
\end{center}
\end{table*}

The atmospheric parameters of the objects used in the comparison between the model atmosphere codes {\sc cmfgen} and {\sc tlusty} are listed in Table \ref{tab_tca}. Table \ref{tab_atparam} contains the atmospheric parameters of the additional objects which have been analysed solely with {\sc tlusty}.

\citet{cro06} used the non-LTE model atmosphere code {\sc cmfgen} to deduce the atmospheric parameters listed in Table \ref{tab_tca}.
They estimated surface gravities at individual temperatures for a second order log~{\em g} -- $T_{\rm eff}$ fit based on previous studies of B-type supergiants. An initial $\xi$ = 20 km\,s$^{-1}$ was assumed, with values considered in the range of 10 -- 40 km\,s$^{-1}$ in multiples of 5 km\,s$^{-1}$, if the fits of helium and silicon lines were not in agreement. Values of the projected rotational velocity ($\upsilon$\,sin{\it i}) were taken from \citet{how97}, with the exception of HD\,14956 which was not included in their sample. For their temperature estimate, \citet{cro06} employed the silicon ionization balance with lines of Si\,{\sc ii} (4128 and 4130 \AA), Si\,{\sc iii} (4552, 4568 and 4574 \AA) and Si\,{\sc iv} (4089 \AA). The Si\,{\sc iii} and Si\,{\sc iv} lines were used for B0.7 -- B2 supergiants, and Si\,{\sc ii}/Si\,{\sc iii} for B2.5 -- B3 objects.

The {\sc tlusty} atmospheric parameters have been determined using an interrelated iterative process (see \citealt{{tho06},{tho07}}). 
Effective temperatures, $T_{\rm eff}$, were determined using the silicon ionization balance from the multiplets of Si\,{\sc ii} (4128, 4130 \AA) and Si\,{\sc iii} (4552, 4567, 4574 \AA) and the Si\,{\sc iv} line at 4116 \AA\ (also the 4088 \AA\ Si\,{\sc iv} line for HD\,126341). Similarly to \citet{cro06}, the Si\,{\sc iii}/Si\,{\sc iv} lines were employed for stars in the spectral range B0.5--B1.5, while Si\,{\sc ii}/Si\,{\sc iii} lines were used for B2--B7 objects. However, \citet{cro06} employed the Si\,{\sc iv} line at 4088 \AA\ (which can be blended with O\,{\sc ii} 4089.29 \AA) rather than 4116 \AA, as used here. In the B1.5 and B2 stars, in addition to other silicon lines, blended or weak features of both Si\,{\sc ii} and Si\,{\sc iv} were observed and the use of these lines is discussed below. 

Surface gravity estimates, log~{\em g}, were obtained by over-plotting theoretical profiles on to the observed spectra in the region of the Balmer lines, with an associated error of $\pm$0.2 dex due to the uncertainty in profile fitting. Note that only the Balmer lines H$\delta$ and H$\gamma$ have been used, as wind effects appear small for these lines but can be significant for H$\alpha$ and H$\beta$ \citep{mce99}.
The microturbulence, $\xi$, was determined using the Si\,{\sc iii} multiplet around $\sim$ 4560 \AA, by plotting abundance against line strength and determining when the gradient is zero. O\,{\sc ii} lines were not used as microturbulence estimates from this species yields larger values than those from Si\,{\sc iii} lines, especially in the case of supergiants \citep{{mce99},{tru04}}. Values of $\xi$ in Tables \ref{tab_tca} and \ref{tab_atparam} have uncertainties of at least 5 km s$^{-1}$. 

The effect of rotational line broadening, $\upsilon$\,sin{\it i}, was calculated by overlaying instrumentally broadened theoretical profiles, with the appropriate atmospheric parameters, on to the individual lines of the Si\,{\sc iii} multiplet. Theoretical profiles were scaled to the same equivalent width as the observed features, convolved to the same spectral resolution, and then rotationally broadened and overlain on to the spectra, to determine the most appropriate $\upsilon$\,sin{\it i} value (see \citealt{hun07} for further details).  
Other large-scale forms of broadening, such as macroturbulence (see for example \citealt{rya02}), were not considered due to the difficulty in distinguishing between  types of broadening through profile fitting. Thus, the derived $\upsilon$\,sin{\it i} values should be considered as upper limits. With the exception of HD\,53138, these estimates agreed with those of \citet{how97} and \citet{cro06} to within $\pm$ 6 km\,s$^{-1}$.

As stated above, in the B1.5 and B2 stars, blended or weak features were observed for Si\,{\sc ii} and Si\,{\sc iv}. 
In the cases of HD\,126341, HD\,14956, HD\,142758 and HD\,141318, the Si\,{\sc ii}/Si\,{\sc iii} ionization balance gave a $T_{\rm eff}$ estimate approximately 900 K larger than obtained using the Si\,{\sc iii}/Si\,{\sc iv} lines, with the gravity estimate changing by $\sim$ 0.1 dex. As all of the features appear reliable, intermediate values of $T_{\rm eff}$ and log~{\em g} were considered the most appropriate for the objects and these are listed in Tables \ref{tab_tca} and \ref{tab_atparam}.
For the other B2 objects (HD\,14818 and HD\,14143), the Si\,{\sc ii} lines were used, as the Si\,{\sc iv} features appeared weak and unreliable and thus were not employed to estimate the effective temperature. 
 
The atmospheric parameters of HD\,152235, HD\,154090, HD\,14956, HD\,14818 and HD\,14143 lie at or below the lowest available gravity point in the metal line grid. However, using the method detailed above, it was possible to deduce atmospheric parameters by extrapolating the silicon abundance estimates, and fitting the Balmer lines using the higher resolution grid. 
This was not believed to be a significant source of error, as tests performed using the available grid points indicated a difference of $<$ 0.1 dex between silicon abundances determined using adjacent grid points.

\begin{figure}
\includegraphics[angle=0,width=0.5\textwidth]{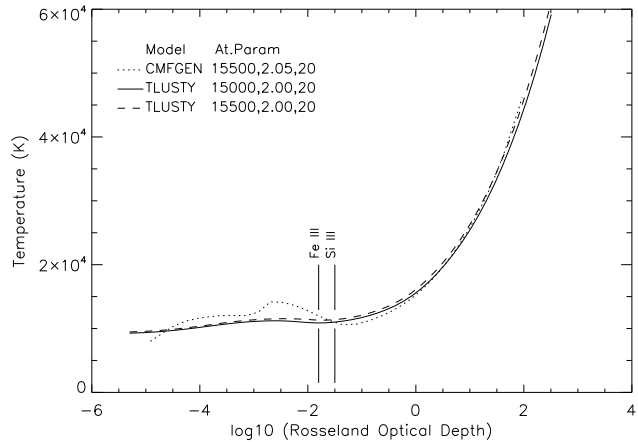}
\caption{Temperature structure plotted 
against Rosseland mean optical depth for {\sc cmfgen} and {\sc tlusty}, with the curves representing the different atmospheric parameters (At.param: $T_{\rm eff}$, log~{\em g} and $\xi$) for HD\,53138, as calculated using the two models. The {\sc tlusty} curves are for the closest available grid points to the derived parameters of $T_{\rm eff}$ = 15200 K and $\xi$ = 18 km\,s$^{-1}$. The regions of formation for Fe\,{\sc iii} and Si\,{\sc iii} are indicated.}
\label{fig_temp}
\end{figure}

In general, good agreement is found between the atmospheric parameters of the two studies, within the errors, despite the use of different model atmospheres. Comparing the temperature structures of {\sc cmfgen} and {\sc tlusty} as a function of the Rosseland optical depth for HD\,53138 (Fig. \ref{fig_temp}), indicates that the temperature structures are in good agreement in the regions of formation for Fe\,{\sc iii} and Si\,{\sc iii}, i.e. log$_{10}$(Rosseland optical depth) $\sim$ --1.8 and --1.5, respectively.  
However, there are differences in the temperature and gravity estimates for HD\,148688, HD\,14956 and HD\,14143. 
This may be due to the use of different silicon lines, especially as the {\sc cmfgen} and {\sc tlusty} atmospheric parameters were deduced from the same observational data for the latter two objects. 
The $\upsilon$\,sin{\it i} values found here tend to be lower than those of \citet{cro06}, with the exception of HD\,154090. In general, the $\upsilon$\,sin{\it i} estimates derived here are thought to be more reliable due to the improved observational data and methods of analysis.

\subsection{Iron line selection and abundances}
\label{sec_femethod}

\begin{table*}
\begin{minipage}{\textwidth}
\caption{Sample of the Fe\,{\sc iii} wavelengths and transitions, with atomic data taken from the Kurucz database$^a$ (http://nova.astro.umd.edu, \citealt{{hub88},{hub95},{hub98}}) and \citet{nah96}$^b$. Full details available on-line.} 
\label{tab_atdata}
{\scriptsize
\begin{tabular}{@{}cccccccccccccc}
\hline
Wavelength&Transition$^b$    		&g$_{i}$$^b$	&g$_{j}$$^b$	& Kurucz$^{a}$	&N$\&$P$^{b}$	&&Wavelength&Transition$^b$    		&g$_{i}$$^b$	&g$_{j}$$^b$	& Kurucz$^{a}$	&N$\&$P$^{b}$	\\
(\AA)	&				&		&		& log\,{\em gf} & log\,{\em gf}	&&(\AA)	&				&		&		& log\,{\em gf} & log\,{\em gf}	  \\
\hline
3953.76 &d$^{3}$G$^{e}$--w$^{3}$G$^{o}$	& 9	& 9	&-1.834  & -2.153	&&4365.64 &a$^{5}$P$^{e}$--z$^{5}$P$^{o}$ &3	 &3	 &-3.404  & -3.305	 \\
																				 \\
4005.04 &e$^{3}$F$^{e}$--z$^{3}$F$^{o}$	& 9	& 9	&-1.755  & -1.810	&&4371.34 &a$^{5}$P$^{e}$--z$^{5}$P$^{o}$ &7	 &5	 &-2.992  & -2.813	 \\
																				 \\
4022.11 &c$^{5}$F$^{e}$--t$^{5}$F$^{o}$	& 9	& 9	&-3.268  & -3.438	&&4382.51 &a$^{5}$P$^{e}$--z$^{5}$P$^{o}$ &5	 &5	 &-3.018  & -3.562	 \\
4022.35 &e$^{3}$F$^{e}$--z$^{3}$F$^{o}$	& 7	& 7	&-2.054  & -1.967											 \\
	&				&	&	&	 &		&&4419.60 &a$^{5}$P$^{e}$--z$^{5}$P$^{o}$ &7	 &7	 &-1.690  & -2.516	 \\	 
4035.43 &y$^{7}$P$^{o}$--c$^{7}$S$^{e}$	& 5	& 7	& 0.119  &  0.147											 \\
	&				&	&	&	 &		&&4431.02 &a$^{5}$P$^{e}$--z$^{5}$P$^{o}$ &5	 &7	 &-2.572  & -2.819	 \\
4039.16 &e$^{3}$F$^{e}$--z$^{3}$F$^{o}$	& 5	& 5	&-2.349  & -2.091											 \\
	&				&	&	&	 &		&&4569.76 &d$^{1}$G$^{e}$--y$^{1}$G$^{o}$ &9	 & 9	 &-1.870  & -1.701	 \\
4053.11 &y$^{7}$P$^{o}$--c$^{7}$S$^{e}$	& 7	& 7	& 0.261  &  0.291											 \\
4053.47 &u$^{5}$F$^{o}$--g$^{5}$D$^{e}$	& 3	& 5	&-1.439  & -1.273	&&5063.42 &b$^{5}$D$^{e}$--z$^{5}$P$^{o}$ & 1	 & 3	 &-2.950  & -3.087	 \\
\hline
\end{tabular}			 
}
\end{minipage}
\end{table*}

Fe\,{\sc iii} lines were identified from synthetic spectra produced with the {\sc tlusty} line list, using a similar method to that discussed by \citet{gri96}. Spectra were computed covering the wavelength range observed with the FEROS data, with atmospheric parameters corresponding to a typical B-type star of spectral type B2, e.g. HD\,141318, as this spectral type appears to have the strongest Fe\,{\sc iii} spectrum from our sample. (Our models indicated that Fe\,{\sc iii} peaks at $T_{\rm eff}$ $\sim$ 21000 K). 
Two sets of abundances were considered; one with a solar abundance for all elements, and the other with all elements excluded apart from iron. 
These were then overplotted on to the HD\,141318 spectra and examined for Fe\,{\sc iii} features. This process was then repeated for the B3 object HD\,79447 to confirm the selection. This object was chosen despite having a weaker Fe\,{\sc iii} spectrum than the B2 objects, as it is a slower rotator and hence possible blended features could be more easily identified. 

The amount of rotational line broadening in the theoretical spectra was increased to observe the effects on the identified lines. If two or more iron features could not be resolved, these were treated as blends within {\sc tlusty}. Where an Fe\,{\sc iii} line was blended with an element of a different species, e.g. O\,{\sc ii}, the iron line was included provided that the features could be adequately resolved from each other. A number of Fe\,{\sc iii} lines that have been considered in previous studies were excluded from our investigation if they were deemed too weak or blended with other species. In particular, we note the following:

\begin{itemize}

\item 
The feature at 4395 \AA, detected by \citet{{van72a},{pin93}} and \citet{zon98}, was found to be blended with O\,{\sc ii} at 4395.93 \AA. 
\item 
A Fe\,{\sc iii} line at 4168.40 \AA, included by \citet{pin93}, is a blend of Fe\,{\sc iii} (4168.45 \AA) and S\,{\sc ii} (4168.38 \AA). 
\item
For the Fe\,{\sc iii} lines around 4372 \AA\ it was not possible to resolve the individual features in the {\sc tlusty} models, or in the slowest rotating object, HD\,79447. Also, they appear to be blended with C\,{\sc ii} lines.
\item
There are a number of Fe\,{\sc iii} lines in the range $\sim$ 5880 -- 6000 \AA, but this spectral region contains telluric transitions, making identification and measurement difficult. 

\end{itemize}

As noted in Section \ref{sec_model}, all of the {\sc tlusty} (LTE) abundance estimates were generated using the Galactic grid and are given in Tables \ref{tab_tcab}, \ref{tab_sab} and \ref{tab_mab}. The {\sc cmfgen} (non-LTE) abundances (also shown in Table \ref{tab_tcab}) were obtained by calculating synthetic spectra for each star, using the appropriate atmospheric parameters, for iron abundances of Fe = 7.35, 7.65 and 7.95 dex. Equivalent widths were then measured and compared to the observed values, and abundances obtained by interpolation or extrapolation. As for the {\sc tlusty} models, this should normally not lead to significant errors. 
In Table \ref{tab_tcab}, the absolute abundance estimates from the {\sc tlusty} models, for HD\,152235, HD\,154090, HD\,14956, HD\,14818 and HD\,14143, have been extrapolated, as the atmospheric parameters lie at or below the lowest available gravity grid points.  
However, tests using the available grid points indicate that this method yields reliable abundance estimates and should not lead to significant errors. 

The abundance values have associated errors due to both systematic and random errors (see \citealt{hun05} for further details). For the {\sc tlusty} comparisons, systematic errors arise due to the uncertainties in the atmospheric parameters, and are estimated by varying the parameters by their associated errors. These error estimates have also been adopted for the {\sc cmfgen} comparison. Random errors are related to the data analysis, such as the oscillator strengths, uncertainties in observations and line fitting errors. Thus, the random uncertainty is assumed to be the standard deviation of the abundance estimates from individual features, divided by the square root of the number of lines observed for that species. The total uncertainty is then taken as the square root of the sum of the squares of the systematic and random errors.

\section{Results and Discussion}
\label{sec_discuss}

\subsection{Comparison of {\sc cmfgen} and {\sc tlusty} results}

\begin{figure}
\includegraphics[angle=0,width=0.5\textwidth]{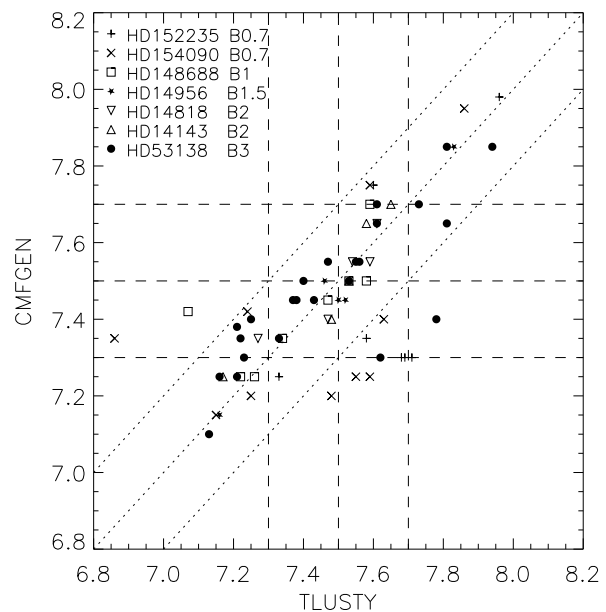}
\caption{Abundance estimates from {\sc tlusty} plotted against those from 
{\sc cmfgen} for Fe\,{\sc iii} features in all the comparison supergiants. 
The ($\cdots$) lines represent the x = y (45${^\circ}$) relation (plus offsets of $\pm$ 0.2 dex), 
while the (-\,-\,-) lines represent x = 7.5 dex and y = 7.5 dex (plus $\pm$ 0.2 dex offsets).}
\label{fig_ct}
\end{figure}

As discussed in Section \ref{sec_model}, there are different physical assumptions in the model atmosphere codes {\sc cmfgen} and {\sc tlusty}. In particular, {\sc tlusty} assumes plane-parallel geometry, while {\sc cmfgen} uses spherical geometry and considers wind properties. Additionally, non-LTE effects have been included in the {\sc cmfgen} Fe\,{\sc iii} calculations.
The effects of the different assumptions on Fe\,{\sc iii} are considered for the sample of supergiants, shown in Table \ref{tab_tca}, which includes the atmospheric parameters determined here using {\sc tlusty} and from \citet{cro06}.
Table \ref{tab_tcab} shows the absolute abundance estimates deduced using the two model atmosphere codes, with the average values calculated from all of the observed Fe\,{\sc iii} features, with the exception of those at 4053 and 4222 \AA. In the {\sc cmfgen} models with higher temperature and lower iron abundance, these features go into emission, and thus no abundance estimates could be obtained for the B0.7 -- B1.5 objects. To maintain consistency, these lines have also been excluded in the B2 -- B3 stars, but this does not significantly affect the average values.

In general, the average abundance estimates are in excellent agreement, having a typical difference of less than 0.05 dex, with HD\,152235 displaying a larger discrepancy of 0.18 dex. Differences in the abundance estimates for the individual lines are generally less than 0.2 dex, although there are some exceptions. 
Fig. \ref{fig_ct} shows the abundance estimates from {\sc tlusty} and {\sc cmfgen} for each Fe\,{\sc iii} transition observed in all the supergiants. 
Assuming that all the stars have the same iron abundance, these values would be expected to be concentrated around a single point. To investigate possible sources of error two approaches have been adopted. Firstly, if errors due to the use of different physical assumptions predominate, abundance estimates would be transposed either horizontally or vertically. Given the simplification adopted in the {\sc tlusty} calculations, horizontal shifts would appear more probable. Alternatively, if errors were present in both analyses (for example, due to incorrect oscillator strengths), points would be shifted along a 45${^\circ}$ line. To illustrate the former, horizontal and vertical loci at an abundance of 7.5 dex (typical of the Galactic value) plus offsets by $\pm$0.2 dex are illustrated, while for the latter, loci at 45$^{\circ}$ plus offsets by $\pm$0.2 dex are shown.

From Fig. \ref{fig_ct}, the majority of points appear to lie along the 45${^\circ}$ line. However, there are some estimates which fall outside of this relation. These are due to the 5833 \AA\ line (in HD\,154090 and HD\,148688), and most of the features in the spectral range 4419--5193 \AA\ 
observed in the two B0.7 stars (HD\,152235 and HD\,154090). This suggests that the differences in physical assumptions may become more significant at higher temperatures and longer wavelengths. High abundance estimates situated along the 45${^\circ}$ relation include the 4273 and 4296 \AA\ features, which are from the same multiplet, suggesting that the oscillator strengths of this multiplet may be too low. Conversely, estimates due to the 4005 \AA\ line for almost all of the supergiants are low, strongly suggesting that the adopted oscillator strength of this transition is too high. There are also other anomalous points, due to lines at 4164, 4166, 4352, 5193, 5306 and 5573 \AA. These features, when observed in other objects, have abundance estimates situated between the horizontal loci, thus making it difficult to confirm their behaviour.

The Fe ionization structure included in the model atmosphere codes {\sc cmfgen} and {\sc tlusty} may not be correct, potentially leading to erroneous estimates of the relative strengths of Fe\,{\sc ii}, Fe\,{\sc iii} and Fe\,{\sc iv} lines. Examining the Fe ionization structure in {\sc cmfgen}, in cooler stars (e.g. HD\,53138) Fe$^{2+}$ is found to be dominant in the region of Fe\,{\sc iii} line formation. However, in the hotter stars (e.g. HD\,152235) Fe$^{2+}$ and Fe$^{3+}$ compete for dominance in the Fe\,{\sc iii} line formation region.
Incorrect ionization structure should lead to a similar shift observed along the 45${^\circ}$ line for all the Fe\,{\sc iii} lines. However, such a trend is not observed in the sample of supergiants, suggesting that uncertainties in the ionization structure are not a significant source of error.

Therefore, the major sources of error appear to be present in both analyses, as the points are distributed along the 45${^\circ}$ line. The most likely causes are observational uncertainties, such as equivalent width measurements, due to blending and poor quality data (especially in the later 
spectral type objects), or incorrect log\,{\it gf} values. Differences between the average abundances obtained from the two approaches are relatively minor, as there are few points lying outside the 45${^\circ}$ relation. Thus, it appears that in general the different physical assumptions, in particular non-LTE effects, are small and a LTE approximation may be a reliable assumption for the Fe\,{\sc iii} spectrum in B-type stars.

\subsection{Assessment of individual Fe\,{\sc iii} lines}
\label{sec_quality}

Although the mean stellar abundance estimates agree, to within the errors, with the Galactic metallicity value, the reliability of the individual features varies. The majority of the Fe\,{\sc iii} features observed are due to single lines or blends of lines from the same multiplet, but some are blends with other multiplets (see Table \ref{tab_atdata}). The reliability of individual lines is discussed below in terms of equivalent width measurements and abundance estimates (see Tables \ref{tab_sab} and \ref{tab_mab}).

\begin{itemize}

\item {\it 3953 \AA}: 
A weak feature only observed in the narrowest-lined stars (HD\,159110, HD\,126341 and HD\,79447). Abundance estimates for this line are in excellent agreement with the average for these objects.

\item {\it 4005, 4022, 4039 \AA}: 
The 4005 \AA\ line is a relatively weak feature and is situated in the wings of He\,{\sc i} 4009 \AA. Its abundance estimate is consistently lower than the average for the individual objects by more than 0.15 dex. However, the other two features (4022 and 4039 \AA) yield similar abundances, indicating that the values for this multiplet are consistent, although there may be problems with the oscillator strengths. The 4022 \AA\ feature in the supergiant HD\,159110 gives a larger abundance than the other features observed, by $\sim$ 0.4 dex, suggesting it may not be reliable. 

\item {\it 4035, 4053, 4081 \AA}: 
The 4035 \AA\ line observed in HD\,79447 yields a larger abundance than the other features of this multiplet. This may be due to blending with N\,{\sc ii} 4035.10 \AA\ and O\,{\sc ii} 4035.49 \AA, making the equivalent width measurement unreliable. The 4053 and 4081 \AA\ features are in good agreement, despite 4053 \AA\ being a blend of two lines. In general, their abundance estimates are lower than the average value obtained, and thus may not be reliable.

\item {\it 4122 -- 4166 \AA}: 
The 4164.92 \AA\ feature is close to S\,{\sc ii} 4165.10 \AA. However, this does not appear to be a problem, as the abundance estimate from the line agrees with that from other features in the same multiplet. The 4166 \AA\ line in HD\,51309 provides a low abundance as it is a weak feature. All lines apart from 4154 \AA\ are blends but as the abundance values are consistent, this is not considered a significant source of error. 

\item {\it 4222 -- 4261 \AA}: 
The 4222 \AA\ feature consistently gives a larger abundance than the average for a given star, which may be due to this line being treated as a blend within {\sc tlusty} and the difficulties in accurately determining the equivalent width, potentially leading to an overestimation of its value. Other features 
from the same multiplet are observed in HD\,159110, HD\,126341 and HD\,79447. In the former two stars, the abundance estimates from these lines tend to be lower than that of 4222 \AA, but in HD\,79447 the 4248 \AA\ line is in agreement with 4222 \AA, suggesting the value may be reliable. The 4248 and 4261 \AA\ lines are consistent for HD\,79447 but not HD\,159110. However the lines are extremely weak in the supergiant and hence have large uncertainties in their equivalent width measurements. 

\item {\it 4273 -- 4310 \AA}: 
These features are all components of the same multiplet. In general they tend to give abundance estimates higher than the overall average for the stars in which they are observed. In the two main-sequence stars, these lines agree well, while in the supergiants 4273 \AA\ tends to yield a larger value than both 
the average abundance estimate and those from the other lines of the multiplet. This may be due to the line being close to O\,{\sc ii} 4273.10 \AA, and thus it is difficult to resolve in stars with broader features.

\clearpage
\begin{landscape}
\begin{table}
\caption{Fe\,{\sc iii} equivalent width measurements (EW)$^{\star}$ and absolute abundance estimates$^\diamond$ for the supergiants derived using {\sc tlusty} (T; LTE) and {\sc cmfgen} (C; non-LTE).}
\label{tab_tcab}
{\scriptsize
\begin{tabular}{@{\extracolsep{0.11cm}}@{}lcccccccccccccccccccccccc}
\hline
		&\multicolumn{3}{c}{HD\,152235}
						&\multicolumn{3}{c}{HD\,154090}
					                     	    	    	&\multicolumn{3}{c}{HD\,148688}
					              	     	    	    				       	&\multicolumn{3}{c}{HD\,14956}
					              	     	    	    				        			       	&\multicolumn{3}{c}{HD\,14818}
				        	      	     	    	    				        							       	&\multicolumn{3}{c}{HD\,14143}
				              		     	    	    				        											       	&\multicolumn{3}{c}{HD\,53138}  \\	 
		\cline{2-4}			\cline{5-7}			\cline{8-10}			\cline{11-13}			\cline{14-16}			\cline{17-19}			\cline{20-22}			\\[0.5mm] 	  
Wavelength	& EW	 & T$^\dagger$  & C  	& EW	 & T$^\dagger$& C  	& EW	 & T	   & C  	& EW	 & T$^\dagger$ & C  	& EW	 & T$^\dagger$  & C  	& EW	 & T$^\dagger$  & C  	& EW	 & T	   & C  	\\	  	  
(\AA)		& (m\AA) & 	   &		& (m\AA)   & 	   &		& (m\AA)   &	   &		& (m\AA)   & 	   &		& (m\AA) & 	   &		& (m\AA) & 	   &		& (m\AA) &	   &		\\	  	 
\hline
4005.04		&17 c    &7.33    &7.25			&14 c    &7.15    &7.15			&24 c    &7.22    &7.25			&28 c	  &7.16     &7.15     		&37 c 	 &7.27    &7.35	   	&30 c	 &7.17    &7.25	   	&25 b 	 &7.23 	  &7.30 	\\	
4053.00$^a$	&20 c	 &7.56	  &$\times$		&$\cdots$&$\cdots$&$\cdots$     	&$\cdots$&$\cdots$&$\cdots$     	&$\cdots$ &$\cdots$ &$\cdots$		&$\cdots$&$\cdots$&$\cdots$ 	&$\cdots$&$\cdots$&$\cdots$	& 7 c  	 &7.24 	  &7.35 	\\	   
4164.00$^a$	&$\cdots$&$\cdots$&$\cdots$		&$\cdots$&$\cdots$&$\cdots$		&$\cdots$&$\cdots$&$\cdots$		&$\cdots$ &$\cdots$ &$\cdots$		&$\cdots$&$\cdots$&$\cdots$	&$\cdots$&$\cdots$&$\cdots$	&23 c	 &7.22    &7.35 	\\	
4166.00$^a$	&$\cdots$&$\cdots$&$\cdots$		&$\cdots$&$\cdots$&$\cdots$		&$\cdots$&$\cdots$&$\cdots$		&$\cdots$ &$\cdots$ &$\cdots$		&$\cdots$&$\cdots$&$\cdots$	&$\cdots$&$\cdots$&$\cdots$	& 8 c	 &7.21    &7.25 	\\	
4222.00$^a$	&17 c	 &7.67	  &$\times$		&19 c	 &7.64	  &$\times$		&23 c	 &7.75	  &$\times$		&19 c	  &7.56	    &$\times$		&24 b	 &7.70	  &7.95		&26 c	 &7.74	  &7.81		&18 b 	 &7.97	  &8.15		\\	
4273.00$^a$	&16 c	 &7.96	  &7.98	       		&16 c	 &7.86	  &7.95	       		&$\cdots$&$\cdots$&$\cdots$     	&16 c	  &7.83     &7.85     		&10 c 	 &7.61    &7.65	    	&11 c	 &7.65    &7.77	   	& 8 c  	 &7.94 	  &7.85 	\\	   
4296.00$^a$	&14 c	 &7.60	  &7.75	       		&16 c	 &7.59	  &7.75	       		&15 c	 &7.59	  &7.70	        	&$\cdots$ &$\cdots$ &$\cdots$		&$\cdots$&$\cdots$&$\cdots$ 	&16 c	 &7.58    &7.65	   	&10 c 	 &7.81 	  &7.65 	\\	   
4310.00$^a$	&$\cdots$&$\cdots$&$\cdots$     	&11 c	 &7.24	  &7.42	       		&$\cdots$&$\cdots$&$\cdots$     	&17 c	  &7.46     &7.50     		&19 c 	 &7.54    &7.55	    	&$\cdots$&$\cdots$&$\cdots$	& 9 c  	 &7.56 	  &7.55 	\\	   
4352.58		&$\cdots$&$\cdots$&$\cdots$		&$\cdots$&$\cdots$&$\cdots$		&$\cdots$&$\cdots$&$\cdots$		&$\cdots$ &$\cdots$ &$\cdots$		&$\cdots$&$\cdots$&$\cdots$	&$\cdots$&$\cdots$&$\cdots$	&23 c	 &7.13    &7.10 	\\	
4382.51		&$\cdots$&$\cdots$&$\cdots$		&$\cdots$&$\cdots$&$\cdots$		&$\cdots$&$\cdots$&$\cdots$		&$\cdots$ &$\cdots$ &$\cdots$		&$\cdots$&$\cdots$&$\cdots$	&$\cdots$&$\cdots$&$\cdots$	&21 c	 &7.81    &7.85 	\\	
4419.60		&50 c	 &7.58    &7.35			&45 c	 &7.55    &7.25			&70 c	 &7.58    &7.50			&91 b	  &7.52     &7.45		&87 b	 &7.47    &7.40		&87 b	 &7.48    &7.40		&86 a	 &7.55    &7.55 	\\	
4431.02		&$\cdots$&$\cdots$&$\cdots$		&$\cdots$&$\cdots$&$\cdots$		&$\cdots$&$\cdots$&$\cdots$		&51 c	  &7.50     &7.45		&66 c	 &7.59    &7.55		&56 c	 &7.53    &7.50		&59 b	 &7.61    &7.65 	\\	
5063.42		&$\cdots$&$\cdots$&$\cdots$		&$\cdots$&$\cdots$&$\cdots$		&$\cdots$&$\cdots$&$\cdots$		&$\cdots$ &$\cdots$ &$\cdots$		&$\cdots$&$\cdots$&$\cdots$	&$\cdots$&$\cdots$&$\cdots$	&16 c	 &7.38    &7.45 	\\	
5086.00$^a$	&$\cdots$&$\cdots$&$\cdots$		&25 c	 &7.63    &7.40			&29 c	 &7.53    &7.50			&$\cdots$ &$\cdots$ &$\cdots$		&$\cdots$&$\cdots$&$\cdots$	&$\cdots$&$\cdots$&$\cdots$	&41 b	 &7.61    &7.70 	\\	
5127.00$^a$	&69 a	 &7.71    &7.30			&59 c	 &7.48    &7.20			&60 c	 &7.26    &7.25			&$\cdots$ &$\cdots$ &$\cdots$		&$\cdots$&$\cdots$&$\cdots$	&$\cdots$&$\cdots$&$\cdots$	&93 a	 &7.47    &7.55 	\\	
5156.11		&70 b	 &7.69    &7.30			&77 b	 &7.59    &7.25			&73 c	 &7.34    &7.35			&$\cdots$ &$\cdots$ &$\cdots$		&$\cdots$&$\cdots$&$\cdots$	&$\cdots$&$\cdots$&$\cdots$	&90 a	 &7.40    &7.50 	\\	
5193.91		&26 c 	 &7.68    &7.30			&16 c	 &7.25    &7.20			&$\cdots$&$\cdots$&$\cdots$		&$\cdots$ &$\cdots$ &$\cdots$		&$\cdots$&$\cdots$&$\cdots$	&$\cdots$&$\cdots$&$\cdots$	&26 c	 &7.21    &7.38 	\\	
5235.66		&$\cdots$&$\cdots$&$\cdots$		&$\cdots$&$\cdots$&$\cdots$		&$\cdots$&$\cdots$&$\cdots$		&$\cdots$ &$\cdots$ &$\cdots$		&$\cdots$&$\cdots$&$\cdots$	&$\cdots$&$\cdots$&$\cdots$	&26 c	 &7.78    &7.40 	\\	
5272.00$^a$	&$\cdots$&$\cdots$&$\cdots$		&$\cdots$&$\cdots$&$\cdots$		&$\cdots$&$\cdots$&$\cdots$		&$\cdots$ &$\cdots$ &$\cdots$		&$\cdots$&$\cdots$&$\cdots$	&$\cdots$&$\cdots$&$\cdots$	& 9 c	 &7.33    &7.35 	\\	
5276.00$^a$	&$\cdots$&$\cdots$&$\cdots$		&$\cdots$&$\cdots$&$\cdots$		&$\cdots$&$\cdots$&$\cdots$		&$\cdots$ &$\cdots$ &$\cdots$		&$\cdots$&$\cdots$&$\cdots$	&$\cdots$&$\cdots$&$\cdots$	&23 c	 &7.62    &7.30 	\\	
5282.00$^a$	&$\cdots$&$\cdots$&$\cdots$		&$\cdots$&$\cdots$&$\cdots$		&$\cdots$&$\cdots$&$\cdots$		&$\cdots$ &$\cdots$ &$\cdots$		&$\cdots$&$\cdots$&$\cdots$	&$\cdots$&$\cdots$&$\cdots$	&18 c	 &7.37    &7.45 	\\	
5299.93		&$\cdots$&$\cdots$&$\cdots$		&$\cdots$&$\cdots$&$\cdots$		&$\cdots$&$\cdots$&$\cdots$		&$\cdots$ &$\cdots$ &$\cdots$		&$\cdots$&$\cdots$&$\cdots$	&$\cdots$&$\cdots$&$\cdots$	&12 c	 &7.43    &7.45 	\\	
5302.60		&$\cdots$&$\cdots$&$\cdots$		&$\cdots$&$\cdots$&$\cdots$		&$\cdots$&$\cdots$&$\cdots$		&$\cdots$ &$\cdots$ &$\cdots$		&$\cdots$&$\cdots$&$\cdots$	&$\cdots$&$\cdots$&$\cdots$	&15 c	 &7.53    &7.50 	\\	
5306.00$^a$	&$\cdots$&$\cdots$&$\cdots$		&$\cdots$&$\cdots$&$\cdots$		&$\cdots$&$\cdots$&$\cdots$		&$\cdots$ &$\cdots$ &$\cdots$		&$\cdots$&$\cdots$&$\cdots$	&$\cdots$&$\cdots$&$\cdots$	&50 c	 &7.25    &7.40 	\\	
5573.42		&$\cdots$&$\cdots$&$\cdots$     	&$\cdots$&$\cdots$&$\cdots$     	&24 c	 &7.47	  &7.45	        	&$\cdots$ &$\cdots$ &$\cdots$		&$\cdots$&$\cdots$&$\cdots$ 	&$\cdots$&$\cdots$&$\cdots$	& 8 c	 &7.16 	  &7.25		\\ 	   
5833.94		&$\cdots$&$\cdots$&$\cdots$     	&29 c	 &6.86	  &7.35	       		&46 b	 &7.07	  &7.42	        	&$\cdots$ &$\cdots$ &$\cdots$		&$\cdots$&$\cdots$&$\cdots$ 	&$\cdots$&$\cdots$&$\cdots$	&50 a 	 &7.73    &7.70 	\\	   
\hline														     							 	     	 			    		   							
Average$^\ddagger$&	 &7.65    & 7.46  		&	 &7.42    &7.39  		&	 &7.38    &7.43  		&	  &7.49     &7.48     		&	 &7.50    &7.50      	&   	 &7.48    &7.50   	&    	 &7.46	  &7.47 	\\	
$\pm$        	&	 &0.42	  & 0.44		&	 &0.36	  &0.36  		&	 &0.18	  &0.18 		&	  &0.13	    &0.18      		&	 &0.06	  &0.09 	&   	 &0.08	  &0.09 	&    	 &0.18	  &0.18		\\	
\hline
\end{tabular}
}
\medskip\\
$^{\star}$ Equivalent width measurements: `a': accurate to better than $\pm$10 per cent, `b': to better than $\pm$20 per cent, and `c': to less than $\pm$20 per cent.\\
$^\dagger$ Extrapolation of values due to atmospheric parameters being on/below the available {\sc tlusty} grid point.\\ 
$^\diamond$ Logarithmic abundance [M/H] on the scale log[H] = 12.00 (dex).\\
$^a$ Treated as blends in {\sc tlusty}.\\
$^\ddagger$ Calculated using all lines except those at 4053 and 4222 \AA.\\
$\times$ No abundance estimate due to {\sc cmfgen} model appearing to go into emission.
\end{table}
\end{landscape}

\begin{landscape}
\begin{table}
\caption{Fe\,{\sc iii} absolute abundance estimates (Ab.$^\diamond$) and equivalent width measurements (EW)$^{\star}$ for the supergiants.} 
\label{tab_sab}
{\scriptsize
\begin{tabular}{@{}lcc@{\extracolsep{0.11cm}}cccccccccccccccc}
\hline
		&\multicolumn{2}{c}{HD\,155985}
			  		&\multicolumn{2}{c}{HD\,108002}
								&\multicolumn{2}{c}{HD\,142758}
						  					&\multicolumn{2}{c}{HD\,141318}
									  					&\multicolumn{2}{c}{HD\,51309}
																	&\multicolumn{2}{c}{HD\,53138}
																				&\multicolumn{2}{c}{HD\,159110}
											 												&\multicolumn{2}{c}{HD\,164353}
																										&\multicolumn{2}{c}{HD\,91619}\\
		\cline{2-3}		\cline{4-5}		\cline{6-7}		\cline{8-9}		\cline{10-11}		\cline{12-13}		\cline{14-15}		\cline{16-17}		\cline{18-19}		\\[0.5mm]
Wavelength	& EW	 & Ab.   	& EW	 & Ab.		& EW	 & Ab.		& EW	 & Ab.		& EW	 & Ab.		& EW	 & Ab.		& EW   & Ab.		& EW	 & Ab.   	& EW	 & Ab.	    \\
( \AA)    	& (m\AA)& 		& (m\AA)   & 	        & (m\AA) & 	        & (m\AA) & 	        & (m\AA) & 	        & (m\AA) & 	        & (m\AA)& 	        & (m\AA) & 	        & (m\AA) & 	        \\
\hline
3953.76		&$\cdots$&$\cdots$	&$\cdots$&$\cdots$	&$\cdots$&$\cdots$	&$\cdots$&$\cdots$     	&$\cdots$&$\cdots$	&$\cdots$&$\cdots$	& 3 c &7.42 		&$\cdots$&$\cdots$     	&$\cdots$&$\cdots$	 \\
4005.04		&$\cdots$&$\cdots$	&28 c	 &7.18	 	&47 b	 &7.39    	&26 c    &7.25 		&24 c 	 &7.13 		&25 c 	 &7.23 		&18 b &7.14 		&13 c 	 &7.05 		&12 c	 &7.16   	\\
4022.00$^a$	&$\cdots$&$\cdots$	&$\cdots$&$\cdots$	&$\cdots$&$\cdots$	&$\cdots$&$\cdots$     	&$\cdots$&$\cdots$	&$\cdots$&$\cdots$	&10 c &7.60 		&$\cdots$&$\cdots$     	&$\cdots$&$\cdots$	 \\
4039.16		&$\cdots$&$\cdots$	&$\cdots$&$\cdots$	&$\cdots$&$\cdots$	&$\cdots$&$\cdots$     	&$\cdots$&$\cdots$	&$\cdots$&$\cdots$	&10 c &7.21 		& 9 c 	 &7.15     	&$\cdots$&$\cdots$	 \\
4053.00$^a$ 	&$\cdots$&$\cdots$	&$\cdots$&$\cdots$	&$\cdots$&$\cdots$	&17 c	 &7.41	 	&$\cdots$&$\cdots$   	& 7 c 	 &7.24   	&13 b &7.28 		&$\cdots$&$\cdots$     	&$\cdots$&$\cdots$	 \\
4081.01 	&$\cdots$&$\cdots$	&$\cdots$&$\cdots$	&$\cdots$&$\cdots$	&$\cdots$&$\cdots$     	&$\cdots$&$\cdots$   	&$\cdots$&$\cdots$   	&13 c &7.21 		&$\cdots$&$\cdots$     	&$\cdots$&$\cdots$	 \\
4122.00$^a$	&$\cdots$&$\cdots$	&$\cdots$&$\cdots$	&$\cdots$&$\cdots$	&$\cdots$&$\cdots$     	&$\cdots$&$\cdots$   	&$\cdots$&$\cdots$   	&30 c &7.27 		&$\cdots$&$\cdots$     	&$\cdots$&$\cdots$	 \\ 
4137.00$^a$	&$\cdots$&$\cdots$	&$\cdots$&$\cdots$	&$\cdots$&$\cdots$	&$\cdots$&$\cdots$     	&$\cdots$&$\cdots$   	&$\cdots$&$\cdots$   	&24 c &7.21 		&$\cdots$&$\cdots$     	&$\cdots$&$\cdots$	 \\
4139.00$^a$ 	&$\cdots$&$\cdots$	&$\cdots$&$\cdots$	&$\cdots$&$\cdots$	&$\cdots$&$\cdots$     	&$\cdots$&$\cdots$   	&$\cdots$&$\cdots$   	&24 c &7.19 		&$\cdots$&$\cdots$     	&$\cdots$&$\cdots$	 \\
4154.96 	&$\cdots$&$\cdots$	&$\cdots$&$\cdots$	&$\cdots$&$\cdots$	&$\cdots$&$\cdots$     	&$\cdots$&$\cdots$   	&$\cdots$&$\cdots$   	& 8 c &7.40 		&$\cdots$&$\cdots$     	&$\cdots$&$\cdots$	 \\
4164.00$^a$	&$\cdots$&$\cdots$	&$\cdots$&$\cdots$	&53 b 	 &7.23		&62 b 	 &7.49     	&28 c	 &7.13   	&23 c	 &7.22   	&42 c &7.29 		&21 c	 &7.21     	&14 c	 &7.42		\\
4166.00$^a$	&$\cdots$&$\cdots$	&$\cdots$&$\cdots$	&24 c	 &7.36		&29 b	 &7.56     	& 5 c	 &6.84   	& 8 c	 &7.21   	&16 c &7.29 		& 6 c	 &7.23     	& 5 c	 &7.33		\\
4222.00$^a$ 	&20 c	 &7.60		&23 c 	 &7.64 		&56 b	 &8.13		&32 a	 &7.93     	&20 b	 &7.83   	&18 b	 &7.97 		&23 a &7.81 		&10 c	 &7.71 		&10 b	 &8.11	 \\
4238.62 	&$\cdots$&$\cdots$	&$\cdots$&$\cdots$	&$\cdots$&$\cdots$	&$\cdots$&$\cdots$     	&$\cdots$&$\cdots$   	&$\cdots$&$\cdots$   	& 7 c &7.34 		&$\cdots$&$\cdots$     	&$\cdots$&$\cdots$	 \\
4248.00$^a$	&$\cdots$&$\cdots$	&$\cdots$&$\cdots$	&$\cdots$&$\cdots$	&$\cdots$&$\cdots$     	&$\cdots$&$\cdots$	&$\cdots$&$\cdots$	& 5 c &7.35 		&$\cdots$&$\cdots$     	&$\cdots$&$\cdots$	 \\
4261.00$^a$	&$\cdots$&$\cdots$	&$\cdots$&$\cdots$	&$\cdots$&$\cdots$	&$\cdots$&$\cdots$     	&$\cdots$&$\cdots$   	&$\cdots$&$\cdots$   	& 4 c &7.56 		&$\cdots$&$\cdots$     	&$\cdots$&$\cdots$	 \\
4273.00$^a$	&17 c  	 &7.82   	&20 c    &7.91		&24 c	 &8.02		&18 c    &7.95 		&$\cdots$&$\cdots$     	& 8 c	 &7.94		&11 c &7.75 		& 6 c	 &7.89   	&$\cdots$&$\cdots$	 \\
4286.00$^a$	&$\cdots$&$\cdots$	&$\cdots$&$\cdots$	&$\cdots$&$\cdots$	&$\cdots$&$\cdots$	&$\cdots$&$\cdots$ 	&$\cdots$&$\cdots$ 	&12 c &7.65 		&$\cdots$&$\cdots$    	&$\cdots$&$\cdots$	 \\
4296.00$^a$	&16 c  	 &7.53  	&21 c	 &7.64	 	&34 c	 &7.95    	&17 c    &7.66 		&11 c 	 &7.59 		&10 c 	 &7.81 		&11 c &7.52 		&14 c 	 &7.97 		&$\cdots$&$\cdots$	  \\
4304.00$^a$	&$\cdots$&$\cdots$	&$\cdots$&$\cdots$	&$\cdots$&$\cdots$	&$\cdots$&$\cdots$	&$\cdots$&$\cdots$ 	&$\cdots$&$\cdots$ 	&14 c &7.55 		&5 c  	 &7.47 		&$\cdots$&$\cdots$	 \\
4310.00$^a$	&18 c  	 &7.43  	&23 c	 &7.55	 	&39 c	 &7.88    	&30 c    &7.78 		&11 c    &7.47 		&9 c     &7.56 		&16 b &7.56 		&5 c  	 &7.36 		&$\cdots$&$\cdots$	  \\
4352.58 	&$\cdots$&$\cdots$	&$\cdots$&$\cdots$	&$\cdots$&$\cdots$	&$\cdots$&$\cdots$     	&$\cdots$&$\cdots$	&23 c	 &7.13		&15 c &7.14 		&$\cdots$&$\cdots$     	&20 c	 &7.26		\\
4365.64 	&$\cdots$&$\cdots$	&$\cdots$&$\cdots$	&$\cdots$&$\cdots$	&$\cdots$&$\cdots$     	&$\cdots$&$\cdots$	&$\cdots$&$\cdots$	& 7 c &7.22 		&$\cdots$&$\cdots$     	&$\cdots$&$\cdots$	\\
4371.34 	&$\cdots$&$\cdots$	&$\cdots$&$\cdots$	&$\cdots$&$\cdots$	&$\cdots$&$\cdots$     	&$\cdots$&$\cdots$	&$\cdots$&$\cdots$	&18 c &7.18 		&$\cdots$&$\cdots$     	&$\cdots$&$\cdots$	\\
4382.51 	&$\cdots$&$\cdots$	&$\cdots$&$\cdots$	&$\cdots$&$\cdots$	&16 c	 &7.84 		&$\cdots$&$\cdots$	&21 c	 &7.81		&13 c &7.78		&11 c	 &7.67  	&11 c	 &7.69  	\\
4419.60		&41 c  	 &7.51  	&81 b	 &7.52	 	&129 a   &7.68    	&63 a    &7.51  	&81 a	 &7.44          &86 a	 &7.55          &42 a &7.32	        &51 a	 &7.38	        &55 b	 &7.48	       \\
4431.02		&37 c  	 &7.79  	&52 c	 &7.56	 	&89 c	 &7.76    	&43 c    &7.58  	&55 b	 &7.53          &59 b	 &7.61          &27 b &7.40	        &36 c	 &7.52	        &41 c	 &7.62	       \\
5063.42		&$\cdots$&$\cdots$	&$\cdots$&$\cdots$	&32 c	 &7.64    	&12 c    &7.39  	&19 b	 &7.42          &16 c	 &7.38          &11 c &7.39	        &11 c	 &7.45	        &13 c	 &7.50	       \\
5086.00$^a$	&19 c  	 &7.53  	&41 b	 &7.58	 	&65 c	 &7.75    	&34 c    &7.65  	&37 b	 &7.53          &41 b	 &7.61          &21 c &7.47	        &25 b	 &7.58	        &27 c	 &7.67	       \\
5127.00$^a$	&53 b  	 &7.45  	&102 a	 &7.46	 	&139 a   &7.57    	&79 a    &7.50  	&93 a	 &7.40          &93 a	 &7.47          &53 a &7.31	       	&54 b	 &7.35	        &65 b	 &7.49	       \\
5156.11		&62 b  	 &7.49  	&111 a	 &7.49	 	&145 a   &7.55    	&74 a    &7.41  	&93 a	 &7.34          &90 a	 &7.40          &43 a &7.17	       	&46 b	 &7.21	        &56 b	 &7.35	       \\
5193.91 	&$\cdots$&$\cdots$	&29 c	 &7.22	 	&43 c	 &7.36    	&29 c    &7.42  	&31 b	 &7.25          &26 c	 &7.21          &20 b &7.27	       	&21 c	 &7.35	        &21 c	 &7.35	       \\
5235.66 	&27 c  	 &7.42  	&$\cdots$&$\cdots$     	&18 c    &7.35    	&32 c    &7.68  	&30 c	 &7.69          &26 c	 &7.78          &18 b &7.45	       	&$\cdots$&$\cdots$      &$\cdots$&$\cdots$      \\
5272.00$^a$	&$\cdots$&$\cdots$	&$\cdots$&$\cdots$	&38 c	 &7.55	 	&26 c	 &7.56 		&11 c	 &7.21          &10 c	 &7.33          &18 b &7.43	       	&12 c	 &7.64	        &$\cdots$&$\cdots$      \\
5276.00$^a$	&$\cdots$&$\cdots$	&$\cdots$&$\cdots$	&36 c	 &7.40	 	&31 b	 &7.57 		&19 c	 &7.35          &23 c	 &7.62          &21 b &7.41	       	&27 c	 &7.94	        &$\cdots$&$\cdots$      \\
5282.00$^a$	&$\cdots$&$\cdots$	&$\cdots$&$\cdots$	&36 c	 &7.28	 	&42 b	 &7.64 		&20 c	 &7.27          &18 c	 &7.37          &25 a &7.41	       	&9 c	 &7.26	        &7 c	 &7.35	       \\
5284.00$^a$	&$\cdots$&$\cdots$	&$\cdots$&$\cdots$	&11 c	 &7.37		&14 c	 &7.60     	&$\cdots$&$\cdots$	&$\cdots$&$\cdots$	&7 c  &7.36 		&$\cdots$&$\cdots$	&$\cdots$&$\cdots$ 	\\
5299.93 	&$\cdots$&$\cdots$	&$\cdots$&$\cdots$	&23 c	 &7.56	 	&19 c	 &7.50 		&12 c 	 &7.29 		&12 c 	 &7.43 		&13 c &7.31 		&$\cdots$&$\cdots$  	&$\cdots$&$\cdots$     \\
5302.60 	&$\cdots$&$\cdots$	&$\cdots$&$\cdots$	&24 c	 &7.53	 	&22 c	 &7.52 		&12 c	 &7.25 		&15 c	 &7.53 		&14 c &7.33 		&$\cdots$&$\cdots$  	&$\cdots$&$\cdots$     \\
5306.00$^a$	&$\cdots$&$\cdots$	&$\cdots$&$\cdots$	&15 c	 &7.55	 	&21 c	 &7.67 		& 9 c 	 &7.31 		& 6 c 	 &7.25 		&11 c &7.43 		&$\cdots$&$\cdots$  	&$\cdots$&$\cdots$     \\
5573.42		&$\cdots$&$\cdots$	&$\cdots$&$\cdots$	&32 b	 &7.53	 	&10 c	 &7.20 		& 8 c	 &7.10		& 8 c	 &7.16		&8 c  &7.17 		&$\cdots$&$\cdots$  	&$\cdots$&$\cdots$     \\ 
5833.94		&48 b  	 &7.14   	&73 a	 &7.33	 	&134 a   &7.77    	&80 a    &7.80 		&56 a 	 &7.59 		&50 a 	 &7.73 		&46 a &7.52 		&26 b 	 &7.57 		&24 a	 &7.99  	\\
6032.00$^a$	&$\cdots$&$\cdots$     	&$\cdots$&$\cdots$	&$\cdots$&$\cdots$  	&12 c    &6.90		&$\cdots$&$\cdots$	&$\cdots$&$\cdots$	&13 c &7.01 		&$\cdots$&$\cdots$	&$\cdots$&$\cdots$    	\\
6036.00$^a$	&$\cdots$&$\cdots$    	&$\cdots$&$\cdots$	&$\cdots$&$\cdots$  	&12 c	 &7.73		&$\cdots$&$\cdots$	&$\cdots$&$\cdots$	& 6 c &7.48 		&$\cdots$&$\cdots$	&$\cdots$&$\cdots$    	\\ 
\hline      	 		        		      
Average &       &7.52   	&	 &7.51	 	&	 &7.59    	&        &7.56 		&	 &7.36 		&	 &7.48		&     &7.38 		&    	 &7.47 		&        &7.52   	\\
$\pm$	&       &0.23   	&	 &0.11		&	 &0.08    	&        &0.15 		&	 &0.18 		&	 &0.17 		&     &0.16	 	&    	 &0.23 		&        &0.28   	\\
\hline		     	   
\end{tabular}	     	   
}
\medskip\\
$^{\star}$ Equivalent width measurements: `a': accurate to better than $\pm$10 per cent, `b': to better than $\pm$20 per cent, and `c': to less than $\pm$20 per cent.\\
$^\diamond$ Logarithmic abundance [M/H] on the scale log[H] = 12.00 (dex).\\
$^a$ Treated as blends in {\sc tlusty}.\\
\end{table}
\end{landscape}

\begin{table}
\caption{Fe\,{\sc iii} equivalent width measurements (EW)$^{\star}$ and
absolute abundance estimates for the main-sequence stars.}
\label{tab_mab}
{\footnotesize
\begin{tabular}{@{}lccccccccccccccccccccccccccccccccccccccc}
\hline
	 	&\multicolumn{2}{c}{\sc HD\,126341}  &\multicolumn{2}{c}{\sc HD\,79447} \\
Wavelength	& EW 	   & Abundance$^\diamond$	& EW  	   	& Abundance$^\diamond$  \\
( \AA)	 	& (m\AA)   &		  		& (m\AA) 	&		\\
\hline
3953.76		& 4 c	& 7.59		& 3 c	 &   7.68		\\
4005.04		&17 b	& 7.38        	&14 c    &   7.39     \\
4022.00$^a$	&11 c	& 7.39        	& 9 c    &   7.44     \\
4035.43		&$\cdots$&$\cdots$	& 6 c	 &   7.90 		\\
4039.16		&$\cdots$&$\cdots$	& 6 b	 &   7.50	\\
4053.00$^a$ 	&12 b	& 7.43        	& 6 b    &   7.45     \\
4081.01		&14 b	& 7.51        	& 6 b    &   7.38     \\
4122.03$^a$$^b$	&18 b	& 7.55        	&11 c    &   7.73     \\
4122.78$^a$$^b$	&15 b	& $\cdots$    	&12 c    & $\cdots$   \\
4137.00$^a$	&28 b	& 7.61        	&12 b    &   7.39     \\
4139.00$^a$	&29 b	& 7.59        	&13 b    &   7.49     \\
4154.96		& 7 c	& 7.47        	&$\cdots$&$\cdots$    \\ 
4164.73$^a$$^b$	&41 c	& 7.70        	&15 c    &   7.50     \\ 
4164.92$^a$$^b$	&$\cdots$& $\cdots$   	&7  c    & $\cdots$   \\
4166.00$^a$	&15 a	& 7.53        	&8  c    &   7.47     \\
4222.00$^a$	&23 a  	& 8.14		& 8 a	 &   7.80	\\
4238.62		& 6 b	& 7.43      	& 3 c    &   7.56     \\
4248.00$^a$	& 8 c	& 7.56		& 7 c 	 &   8.02 	\\
4261.00$^a$	& 4 c	& 7.57        	& 3 c    &   7.88     \\
4273.00$^a$	& 8 c	& 7.64        	& 5 b    &   7.94     \\ 
4286.00$^a$	&11 b   & 7.73	      	&6  c    &   7.85     \\ 
4296.00$^a$	&13 a   & 7.78        	&7  b    &   7.90     \\
4304.00$^a$	&16 b	& 7.87        	&7  b    &   7.81     \\
4310.00$^a$	&17 b	& 7.88        	&8  b    &   7.80     \\
4352.58 	&10 b	& 7.17        	&10 c    &   7.38     \\
4365.64		&5  c	& 7.24        	&7  c    &   7.55     \\
4371.34		&20 c	& 7.55        	&12 c    &   7.33     \\
4382.51		&11 c	& 7.93        	&11 b    &   8.02     \\
4419.60		&36 a	& 7.77        	&33 a    &   7.59     \\
4431.02		&24 b	& 7.68        	&23 a    &   7.68     \\
4569.76		&$\cdots$&$\cdots$	& 3 c	 &   7.48	\\
5063.42		&11 b	& 7.58        	&10 c    &   7.71     \\
5086.00$^a$	&20 b	& 7.70        	&14 c    &   7.66     \\
5127.00$^a$	&52 a	& 7.67        	&55 b    &   7.77     \\
5156.11 	&40 a	& 7.77        	&32 a    &   7.46     \\
5193.91		&19 b	& 7.49        	&14 c    &   7.55     \\ 
5235.66		&15 b	& 7.60        	&5  b    &   7.49     \\
5272.00$^a$	&9  b	& 7.52        	&6  b    &   7.58     \\
5276.00$^a$	&18 b	& 7.63        	&6  b    &   7.45     \\
5282.00$^a$	&20 a	& 7.60        	&10 b    &   7.59     \\
5284.00$^a$	&6  c	& 7.37        	&5  c    &   7.95     \\
5299.93		&13 b	& 7.51        	&9  b    &   7.83     \\
5302.60		&15 b	& 7.62        	&6  c    &   7.59     \\
5306.00$^a$	&11 c	& 7.58        	&8  c    &   7.91     \\
5375.00$^a$	& 4 c	& 7.49		&$\cdots$&$\cdots$	\\
5460.80		& 9 c	& 7.52		&$\cdots$&$\cdots$	\\
5485.52		& 8 c	& 7.35        	&5  c    &   7.54     \\
5573.42		& 9 b	& 7.28        	& 2 c    &   7.16     \\
5833.94		&36 a	& 7.89        	&15 a    &   7.59     \\
5999.00$^a$	&$\cdots$&$\cdots$     	&10 c    &   7.69     \\
6032.00$^a$	&12 b   & 7.16	      	&6  c    &   7.35     \\
6036.00$^a$	& 5 c   & 7.38 	      	& 2 c    &   7.71	\\
\hline	
Average		&	& 7.57	      	&	 &   7.63     \\
$\pm$		&	& 0.22    	&	 &   0.26 \\
\hline
\end{tabular}
}
\medskip\\
$^{\star}$ EW measurements: `a': accurate to better than $\pm$10 per cent, `b': to better than $\pm$20 per cent, and `c': to less than $\pm$20 per cent.
$^\diamond$ Logarithmic abundance [M/H] on the scale log[H] = 12.00 (dex).
$^a$ Treated as blends in {\sc tlusty}. 
$^b$ Equivalent widths of adjacent lines were combined to produce the corresponding abundances shown.
\\
\end{table}

\item {\it 4352 -- 4431 \AA}: 
None of the lines is a blend. The lines at 4419 and 4431 \AA\ show good agreement for all stars, although in HD\,155985 the 4431 \AA\ abundance is slightly higher, due to the difficulty in resolving this feature in the spectra from N\,{\sc ii} 4431.81 \AA. The 4352 \AA\ line tends to give a lower abundance than the average, while 4382 \AA\ gives a higher value. However, the overall abundance value for this multiplet appears reliable. 
The 4371 \AA\ line is only reliably resolved in the main-sequence objects and HD\,159110 as it is close to a strong blended feature at 4372 \AA\ due to Fe\,{\sc iii} and C\,{\sc ii}. Another component of this multiplet, at 4395.76 \AA, has been excluded as it is blended with O\,{\sc ii} 4395.93 \AA\ (see Section \ref{sec_femethod}).
  
\item {\it 4569 \AA}:
This feature is only observed in HD\,79447 as it is intrinsically weak. The abundance obtained is lower than the average, and does not appear reliable. 

\item {\it 5063 -- 5193 \AA}: 
These are strong, well identified features. The abundance estimates from these lines tend to vary around the average and appear reliable.
The 5086 \AA\ line is a blend of several transitions and gives a slightly larger abundance than the other lines (which are components of the same multiplet) 
for the supergiants, possibly due to poor S/N around the line. As this trend is not observed in the main-sequence stars, the line appears reliable.

\item {\it 5235 -- 5282, 5299, 5302 \AA}: 
For the main-sequence stars there is good agreement between the abundance estimates, but the line at 5299 \AA\ yields a higher value for HD\,79447, and 5276 \AA\ for HD\,164353. The 5276 \AA\ feature is close to Fe\,{\sc ii} 5276.00 \AA, which appears to affect the abundance estimates obtained for HD\,164353. The Fe\,{\sc iii} line at 5235 \AA\ is adjacent to Fe\,{\sc ii} 5234.62 \AA. However, these lines are well resolved in all stars apart from the supergiants of spectral type B3. Thus for these objects (HD\,51309 and HD\,53138) the abundances may be higher due to the difficulty in measuring the individual equivalent widths. The blending of different transitions in the lines does not appear to affect the abundance estimates obtained.

\item {\it 5272, 5284, 5306 \AA}: 
The 5272 and 5306 \AA\ features contain transitions from several multiplets, whereas 5284 \AA\ is a blend lines from the same multiplet. 
Good agreement is found among all the  lines, although the 5284 \AA\ abundance is slightly lower for HD\,126341 and HD\,142758, while the 5272 \AA\ value is slightly lower in HD\,79447. This is probably due to the lines being weak in these stars, but overall these lines appear reliable. 

\item {\it 5375 \AA}:
This is a blend of two transitions, and is only observed in HD\,126341. It has a small equivalent width and it gives a slightly lower than average abundance estimate. 

\item {\it 5460 -- 5573 \AA}: 
The 5573 \AA\ line tends to give a lower than average abundance. The abundances from the other lines are also slightly below the average, but are only observed in the main-sequence stars. In general these are weak lines and thus may not be reliable. 

\item {\it 5833 \AA}:  
The abundance from this feature varies around the average in the stars, despite being a strong, isolated, single transition, and this may be due to non-LTE effects.

\item {\it 5999 \AA}: 
This line is situated close to telluric features (see Section \ref{sec_femethod}) and is only resolved in HD\,79447. However, this abundance estimate appears reliable as it agrees with the overall value for the star. 

\item {\it 6032, 6036 \AA}:  
The 6032 \AA\ line gives a consistently lower than average abundance, while the 6036 \AA\ abundance estimates have a larger scatter. These lines have relatively small equivalent widths, and do not appear reliable due to the low S/N of the region. 

\end{itemize}

As there are a number of Fe\,{\sc iii} lines observed it is possible to compare microturbulence values obtained using the Fe\,{\sc iii} and Si\,{\sc iii} lines. In general, the values obtained are consistent, when using all of the Fe\,{\sc iii} features observed. 
The multiplets of a$^{5}$P$^{e}$--z$^{5}$P$^{o}$ (4352 -- 4431 \AA) and b$^{5}$D$^{e}$--z$^{5}$P$^{o}$ (5063 -- 5193 \AA) could be considered independently and are in good agreement with the overall result, provided well-observed lines are employed.

\subsection{Comparison with previous studies}

\begin{table}
\caption{Summary of previously determined iron abundances from \citet{rei90} (R90), \citet{nie05} (N05), \citet{gie92} (G92) and \citet{pro01} (P01).} 
\label{tab_sum}
\begin{tabular}{@{}lcccccccccccccccc}
\hline
HD	& \multicolumn{2}{c}{Abundance (dex)}	&$\Delta$[Fe/H] & Ref.	\\
Number	& This work	& Previous Work		&		& 	\\
\hline
79447	& 7.63		& 7.62$^{a}$	&+0.01	& R90	\\
	&		& 7.66$^{b}$	&--0.03	& R90	\\
\\	
126341	& 7.57		& 7.11$^{e}$	&+0.46	& N05	\\
\\
51309	& 7.36		& 7.51$^{c}$	&--0.15	& G92	\\
	&		& 7.69$^{d}$	&--0.33	& G92	\\
	& 		& 6.93$^{e}$	&+0.43	& P01	\\
\hline
\end{tabular}
\medskip \\
$^{a}$ Using Fe\,{\sc iii} lines, $^{b}$ Using Fe\,{\sc ii} lines, $^{c}$ $T_{\rm eff}$ = 17390 K, $^{d}$ $T_{\rm eff}$ = 14860 K, $^{e}$ Assuming a Galactic abundance of 7.50 dex, \\
\end{table}

Most of our targets have been analysed previously. For HD\,108002 and HD\,159110, no previous results were found, while for HD\,155985 and HD\,142758 only projected rotational velocities have been determined \citep{{bal75},{how97}}. Only three of the objects have previous iron abundance estimates, namely HD\,79447, HD\,126341 and HD\,51309. These are discussed below and summarized in Table \ref{tab_sum}. 

\citet{rei90} analysed optical data taken with the CASPEC spectrograph on the 3.6-m telescope (ESO) for HD\,79447, covering the 3900 -- 4900 \AA\ wavelength range. They obtained LTE atmospheric parameters of $T_{\rm eff}$ = 17800 K and log~{\em g} = 3.50 dex, and Fe\,{\sc ii} and Fe\,{\sc iii} iron abundances.  Their Fe\,{\sc iii} (and overall iron) abundance is in excellent agreement with that found here. 

In the case of HD\,126341, {\it IUE} spectra spanning 1100 -- 3200 \AA\ were analysed with a best-fit procedure by \citet{nie05}. 
A metallicity was obtained, which is significantly lower than our value, possibly due to the use of UV spectra, as our atmospheric parameters agree well with theirs ($T_{\rm eff}$ $\sim$ 22700 K and log~{\em g} = 3.74 dex). 

There are two previous estimates for the iron abundance in HD\,51309. \citet{gie92} analysed optical data from the McDonald 2.1-m telescope and coud\'{e} spectrograph, identifying two Fe\,{\sc ii} and two Fe\,{\sc iii} lines. For the Fe\,{\sc iii} lines, we found a similar equivalent width to their value for 5063 \AA, but for the 5156 \AA\ feature their result is approximately 50 per cent of that measured here (54 m\AA\ compared to 93 m\AA). Also, they do not observe the line at 5193 \AA, which we have detected, possibly due to the use of higher S/N spectra in the present paper. This was tested by degrading theoretical spectra, calculated using the appropriate atmospheric parameters (from \citealt{gie92}) and Fe abundance for the line (taken from this work), indicating that the feature may have been unclear due to noise. Gies \& Lambert considered two different temperature estimates (17390 and 14860 K) and associated abundances. They state that their data are insufficient to distinguish between these temperatures. However, both temperature estimates are consistent with that found here (16600 K). \citet{pro01} also investigated HD\,51309, using {\it IUE} SWP spectra to obtain an LTE abundance. Their value is much lower than the abundance derived by \citet{gie92} or that found here. As for HD\,126341, this difference is believed to be due to the use of UV spectra rather than optical data, as their effective temperature (16570 K) and surface gravity (2.60 dex) agree well with our estimates. 

It is worth noting that a selection of iron features have been considered in other studies, but with no abundances determined. \citet{len93} observed five of our sample of supergiants (HD\,14956, HD\,14818, HD\,14143, HD\,53138 and HD\,164353). They note that several lines due to Fe\,{\sc iii} are visible (over the range 3950 -- 4950 \AA), but that they are relatively weak and hence unreliable. \citet*{{van72a},{van72b}} investigated HD\,53138 and observed seventeen absorption features due to Fe\,{\sc iii} in their optical (3500 -- 8800 \AA) spectrograms, stating that fourteen are blended with lines of other spectra and giving equivalent width measurements for thirteen. While a number of these lines have been included here, several were excluded as they were considered unreliable due to blending with other features. For example,  Fe\,{\sc iii} 4041.86 \AA\ appears to be blended with O\,{\sc ii} 4041.95 \AA\ (see Section \ref{sec_femethod}). For the lines included here, the equivalent width measurements by \citet{van72a} tend to be larger. Those listed here are believed to be more reliable due to the improved quality of the spectra.

\section{Conclusions}
\label{sec_conc}

Due to their intrinsic weakness, Fe\,{\sc iii} absorption lines have not been widely considered for use in chemical composition studies. Instead, the very rich ultraviolet spectral region has been favored. However, the results found in this paper suggest that the optical region can provide consistent results. The stars here display abundance estimates that agree with the Galactic metallicity, and are consistent with previous studies using optical spectra, where available. By contrast, previous determinations from ultraviolet spectra have followed the trends observed in other studies (for example \citealt{{tho07},{duf07}}), providing lower abundance estimates, in these cases by approximately 0.5 dex.

Although our study has concentrated on B-type stars found in the Milky Way, the optical Fe\,{\sc iii} lines examined here can be applied to studies of B-type stars in other galaxies, provided suitable S/N spectra are employed. For example, \citet{tru02} analysed a sample of B-type supergiants in M\,31, obtaining an Fe abundance from the Fe\,{\sc iii} line at 4419 \AA. \citet{rol02} analysed a sample of OB-type main-sequence stars from the LMC, finding an Fe abundance for one object (PS\,34-16), while \citet{rol03} investigated a B-type dwarf from the SMC (AV\,304), finding agreement with other giant stars. 
More recently, \citet{tru07} used similar methods to those detailed here and obtained Fe abundances for B-type stars in the Galaxy, LMC and SMC, using the two Fe\,{\sc iii} lines at 4419 and 4430 \AA. The results found were consistent with the present accepted metallicities of these systems. These studies indicate that the optical Fe\,{\sc iii} lines can provide reliable abundance indicators in different galaxies. 

Our comparison of stars analysed using the model atmosphere codes {\sc cmfgen} and {\sc tlusty} generally shows little difference in the abundance estimates, indicating that the different physical assumptions, in particular non-LTE effects, are small for this species. Therefore, the results suggest that the optical Fe\,{\sc iii} absorption line spectrum may be used with confidence in chemical composition studies, and an LTE analysis provides reliable results. 

The atomic data of \citet{nah96}, employed both in this paper and by \citet{cro06}, appear to provide appropriate abundances, although there are some features, such as those at 4005 and 4273 \AA, whose log\,{\it gf} values may be incorrect. Comparing the values in Table \ref{tab_atdata} shows that, for some features, e.g. the 4166.88, 4419.60 and 5272.90 \AA\ lines, there are large differences between the atomic data from the Kurucz database and \citet{nah96}. Further work is required to refine the atomic data for this species.

\begin{table}
\caption{Recommended Fe\,{\sc iii} lines for use as abundance diagnostics.} 
\label{tab_touse}
\begin{tabular}{@{}lcccc}
\hline
Line	& Spectral Type		&Line	& Spectral Type	\\
(\AA)	& Range			&(\AA)	& Range		\\
\hline
4419	& B0.5--B7		&5156	& B0.5--B7 	\\	 
4431	& B1--B7		&5272	& B1.5--B5 	\\
5063	& B1.5--B7		&5282 	& B2--B7 	\\
5086	& B0.5--B7		&5299 	& B1.5--B4	\\   	
5127	& B0.5--B7		&5302	& B1.5--B4	\\
\hline
\end{tabular}			 
\end{table}

In Table \ref{tab_touse} we list recommended Fe\,{\sc iii} lines which we believe, based on the present study, will provide reliable diagnostics for the iron abundance in early B-type stars. 
Lines have been selected based on the following criteria:

\begin{itemize}

\item Relatively strong, isolated feature, free from known blends.

\item Yields an abundance estimate within $\pm$ 0.2 dex of the mean iron abundance for all well observed lines.

\end{itemize}

\begin{figure*}
\includegraphics[angle=0,width=0.8\textwidth]{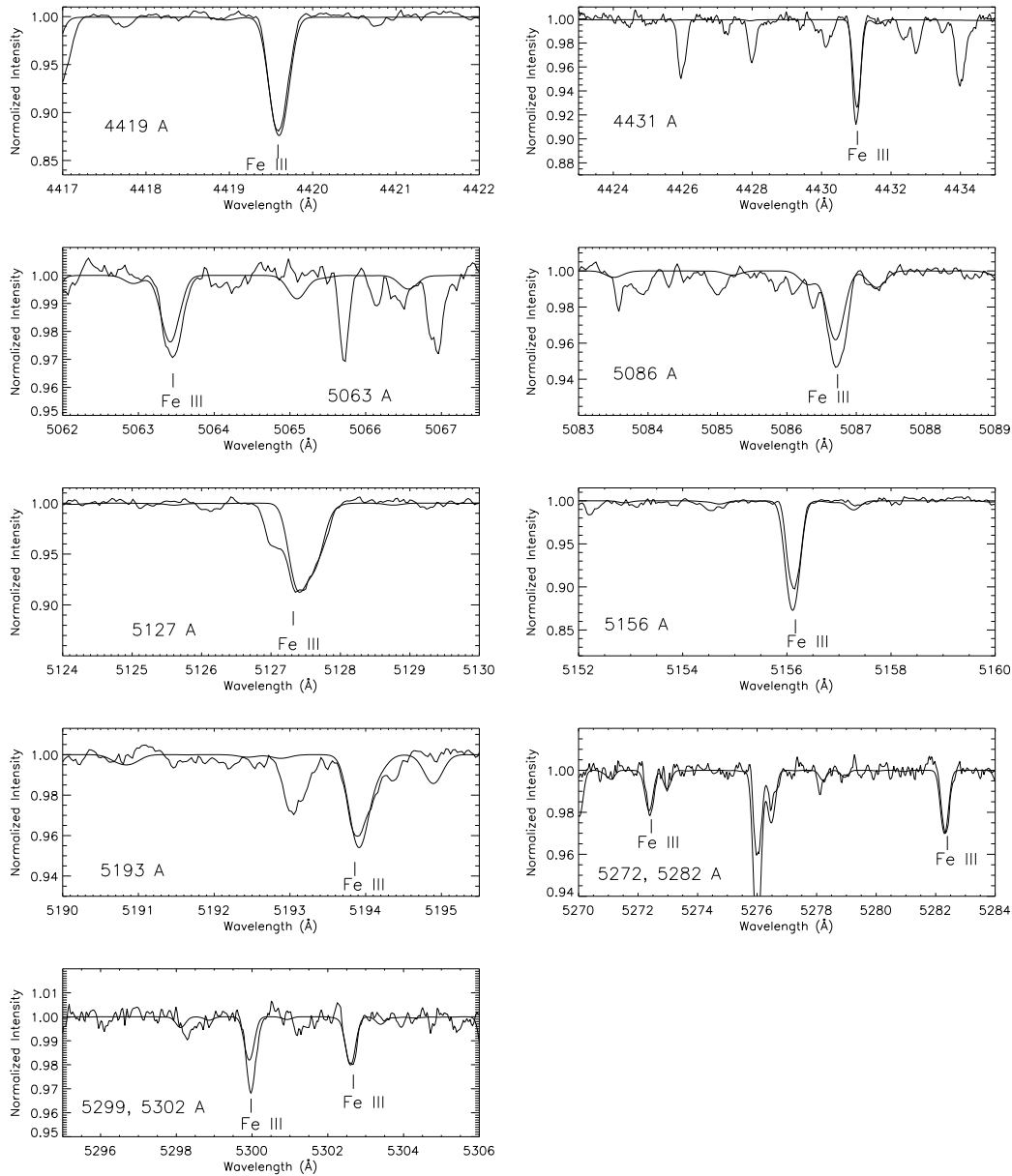}
\caption{Observed Fe\,{\sc iii} lines in HD\,79447 as listed in Table \ref{tab_touse}, including the 5193 \AA\ line. Overplotted are theoretical fits (smooth line) to the Fe\,{\sc iii} lines, calculated using the average abundance estimate for the star of 7.63 dex.}
\label{fig_felines}
\end{figure*}

The range of spectral types over which it is advisable to use the lines as an abundance diagnostic is also listed in Table \ref{tab_touse}. This spectral type range is not the same for all lines, generally due to the fact that some of the weaker transitions are only detected over a restricted span of spectral types. Observations for all of the lines in this Table are shown in Fig. \ref{fig_felines} for HD\,79447. 
The line at 5193 \AA\ has been included in the Figure, but not the Table, because its derived abundance is more than 0.2 dex larger than the mean value for three of the stars studied here, namely HD\,108002 (B1), HD\,142768 (B1.5) and HD\,53138 (B3). However, the feature, along with others observed (see Tables \ref{tab_tcab}, \ref{tab_sab} and \ref{tab_mab}) may be suitable for diagnostic use, depending on the quality of the spectra used. We note that the stars included in this study have been selected due to being narrow lined and are typical of their individual luminosity classes. However, if objects with larger $\upsilon$\,sin{\it i} values were employed, blending between Fe lines and other metal features may occur, thus, care should be taken when using objects with larger projected rotational velocities.

\section*{acknowledgments}
HMAT acknowledges financial support from the Northern Ireland Department for Education and Learning (DEL). 
FPK is grateful to AWE Aldermaston for the award of a William Penney Fellowship. 

{}

\begin{table*}
\setcounter{table}{2}
\begin{minipage}{\textwidth}
\caption{Fe\,{\sc iii} wavelengths and transitions, with atomic data taken from the Kurucz database$^a$ (http://nova.astro.umd.edu, \citealt{{hub88},{hub95},{hub98}}) and \citet{nah96}$^b$.} 
\label{tab_atdata}
{\scriptsize
\begin{tabular}{@{}cccccccccccccc}
\hline
Wavelength&Transition$^b$    		&g$_{i}$$^b$	&g$_{j}$$^b$	& Kurucz$^{a}$	&N$\&$P$^{b}$	&&Wavelength&Transition$^b$    		&g$_{i}$$^b$	&g$_{j}$$^b$	& Kurucz$^{a}$	&N$\&$P$^{b}$	\\
(\AA)	&				&		&		& log\,{\em gf} & log\,{\em gf}	&&(\AA)	&				&		&		& log\,{\em gf} & log\,{\em gf}	  \\
\hline
3953.76 &d$^{3}$G$^{e}$--w$^{3}$G$^{o}$	& 9	& 9	&-1.834  & -2.153	&&4365.64 &a$^{5}$P$^{e}$--z$^{5}$P$^{o}$ &3	 &3	 &-3.404  & -3.305	 \\
																				 \\
4005.04 &e$^{3}$F$^{e}$--z$^{3}$F$^{o}$	& 9	& 9	&-1.755  & -1.810	&&4371.34 &a$^{5}$P$^{e}$--z$^{5}$P$^{o}$ &7	 &5	 &-2.992  & -2.813	 \\
																				 \\
4022.11 &c$^{5}$F$^{e}$--t$^{5}$F$^{o}$	& 9	& 9	&-3.268  & -3.438	&&4382.51 &a$^{5}$P$^{e}$--z$^{5}$P$^{o}$ &5	 &5	 &-3.018  & -3.562	 \\
4022.35 &e$^{3}$F$^{e}$--z$^{3}$F$^{o}$	& 7	& 7	&-2.054  & -1.967											 \\
	&				&	&	&	 &		&&4419.60 &a$^{5}$P$^{e}$--z$^{5}$P$^{o}$ &7	 &7	 &-1.690  & -2.516	 \\	 
4035.43 &y$^{7}$P$^{o}$--c$^{7}$S$^{e}$	& 5	& 7	& 0.119  &  0.147											 \\
	&				&	&	&	 &		&&4431.02 &a$^{5}$P$^{e}$--z$^{5}$P$^{o}$ &5	 &7	 &-2.572  & -2.819	 \\
4039.16 &e$^{3}$F$^{e}$--z$^{3}$F$^{o}$	& 5	& 5	&-2.349  & -2.091											 \\
	&				&	&	&	 &		&&4569.76 &d$^{1}$G$^{e}$--y$^{1}$G$^{o}$ &9	 & 9	 &-1.870  & -1.701	 \\
4053.11 &y$^{7}$P$^{o}$--c$^{7}$S$^{e}$	& 7	& 7	& 0.261  &  0.291											 \\
4053.47 &u$^{5}$F$^{o}$--g$^{5}$D$^{e}$	& 3	& 5	&-1.439  & -1.273	&&5063.42 &b$^{5}$D$^{e}$--z$^{5}$P$^{o}$ & 1	 & 3	 &-2.950  & -3.087	 \\
																				 \\
4081.01 &y$^{7}$P$^{o}$--c$^{7}$S$^{e}$	& 9	& 7	& 0.364  &  0.398	&&5086.52 &w$^{5}$P$^{o}$--e$^{5}$D$^{e}$ & 3	 & 3	 &-1.307  & -2.141	 \\
	&				&	&	&	 &		&&5086.70 &b$^{5}$D$^{e}$--z$^{5}$P$^{o}$ & 5	 & 3	 &-2.590  & -2.845	 \\
4122.03 &y$^{7}$P$^{o}$--b$^{7}$D$^{e}$	& 5	& 5	& 0.436  &  0.431	&&5087.37 &b$^{5}$D$^{e}$--v$^{5}$P$^{o}$ & 5	 & 3	 &-0.733  & -0.778	 \\
4122.78 &y$^{7}$P$^{o}$--b$^{7}$D$^{e}$	&5 	& 3 	& 0.364  &  0.386											 \\
4123.00 &v$^{5}$P$^{o}$--c$^{5}$P$^{e}$ &7 	& 5 	& 0.051  &  0.041      	&&5127.39 &b$^{5}$D$^{e}$--z$^{5}$P$^{o}$ & 7	 & 5	 &-2.218  & -2.423	 \\
	&				&	&	&	 &		&&5127.63 &b$^{5}$D$^{e}$--z$^{5}$P$^{o}$ & 5	 & 5	 &-2.564  & -2.627	 \\
4137.01 &v$^{5}$P$^{o}$--c$^{5}$P$^{e}$	& 7	& 5 	& 0.051  &  0.041        										 \\
4137.13 &y$^{5}$H$^{o}$--b$^{5}$I$^{e}$	& 7	& 9	& 0.719  &  0.808       &&5156.11 &b$^{5}$D$^{e}$--z$^{5}$P$^{o}$ & 9	 & 7	 &-2.018  & -2.140	 \\
4137.76 &y$^{7}$P$^{o}$--b$^{7}$D$^{e}$	& 7	& 9	& 0.644  &  0.658     											 \\				&				&	&	&	 &		&&5193.91 &b$^{5}$D$^{e}$--z$^{5}$P$^{o}$ & 7	 & 7	 &-2.852  & -2.730	 \\
4139.35 &y$^{7}$P$^{o}$--b$^{7}$D$^{e}$	& 7	& 7	& 0.553  &  0.547											 \\
4140.48 &y$^{7}$P$^{o}$--b$^{7}$D$^{e}$	& 7	& 5	& 0.114  &  0.128      	&&5235.66 &a$^{7}$D$^{e}$--y$^{7}$P$^{o}$ & 9	 & 9	 &-0.107  & -0.074	 \\
																				 \\
4154.96 &y$^{7}$P$^{o}$--b$^{7}$D$^{e}$	&11 	&13	& 0.891  &  0.964     	&&5272.37 &a$^{7}$D$^{e}$--y$^{7}$P$^{o}$ & 5	 & 7	 &-0.421  & -0.382	 \\
	&				&	&	&	 &		&&5272.90 &b$^{5}$G$^{e}$--y$^{5}$H$^{o}$ & 9	 & 9	 &-0.409  & -2.229	 \\
4164.73 &y$^{7}$P$^{o}$--b$^{7}$D$^{e}$	& 9	&11	& 0.935  &  0.946     	&&5272.98 &a$^{5}$I$^{e}$--y$^{5}$H$^{o}$ &17	 &15	 & 0.598  &  0.677	 \\
4164.92 &y$^{5}$H$^{o}$--b$^{5}$I$^{e}$	&13	&15	& 1.011  &  1.042											 \\
	&				&	&	&	 &		&&5276.19 &d$^{3}$D$^{e}$--y$^{3}$D$^{o}$ & 5	 & 7	 &-7.667  & -5.334	 \\
4166.84 &y$^{7}$P$^{o}$--b$^{7}$D$^{e}$	&9 	& 9	& 0.409  &  0.431	&&5276.48 &a$^{7}$D$^{e}$--y$^{7}$P$^{o}$ & 7	 & 7	 &-0.001  &  0.036	 \\
4166.88 &y$^{5}$H$^{o}$--h$^{5}$G$^{e}$	&11	& 9	& 0.436  & -0.587											 \\
	&				&	&	&	 &		&&5282.30 &a$^{7}$D$^{e}$--y$^{7}$P$^{o}$ & 9	 & 7	 & 0.108  &  0.146	 \\
4220.12 &v$^{5}$D$^{o}$--g$^{5}$G$^{e}$ & 7 	& 9 	&-0.565  & -0.347      	&&5282.58 &d$^{3}$D$^{e}$--y$^{3}$D$^{o}$ & 7	 & 7	 &-3.648  & -4.432	 \\
4222.27 &w$^{5}$P$^{o}$--d$^{5}$S$^{e}$ & 7 	& 5 	& 0.272  &  0.153     											 \\
	&				&	&	&	 &		&&5284.83 &a$^{5}$I$^{e}$--y$^{5}$H$^{o}$ &13	  &13	  & 0.472  &  0.601	 \\	   
4238.62 &w$^{5}$P$^{o}$--d$^{5}$S$^{e}$	&5 	& 5 	&-0.030  &  0.006	&&5284.83 &a$^{5}$I$^{e}$--y$^{5}$H$^{o}$ &15	  &13	  &-0.836  & -0.531	 \\
																				 \\
4248.34 &v$^{5}$F$^{o}$--h$^{5}$G$^{e}$ &9 	&11 	& 0.288  &  0.099       &&5299.93 &a$^{7}$D$^{e}$--y$^{7}$P$^{o}$ & 3	 & 5	 &-0.166  & -0.129	 \\
4248.77 &w$^{5}$P$^{o}$--d$^{5}$S$^{e}$ &3 	&5 	&-0.095  & -0.217      											 \\
	&				&	&	&	 &		&&5302.60 &a$^{7}$D$^{e}$--y$^{7}$P$^{o}$ & 5	 & 5	 &-0.120  & -0.083	 \\  
4260.31 &v$^{5}$F$^{o}$--h$^{5}$G$^{e}$ &5 	& 7	& 0.326  & -0.168      	 										 \\
4261.39 &z$^{7}$F$^{o}$--a$^{7}$G$^{e}$ &3 	& 5 	& 0.251  & -0.041      	&&5306.13 &a$^{5}$I$^{e}$--y$^{5}$H$^{o}$ & 9	  & 9	  &-0.992  & -0.706	 \\
4261.72 &v$^{5}$F$^{o}$--h$^{5}$G$^{e}$ &5 	& 5	&-0.918  & -0.868      	&&5306.76 &c$^{5}$F$^{e}$--w$^{5}$D$^{o}$ & 7	  & 5	  &-0.341  & -0.306	 \\
	&				&	&	&	 &		 										 \\   
4273.37 &z$^{7}$F$^{o}$--a$^{7}$G$^{e}$ &5 	&5 	& 0.252  & -0.077     	&&5375.47 &d$^{5}$G$^{e}$--u$^{5}$F$^{o}$ &13	  &11	  & 0.461  &  0.543	 \\ 
4273.41 &z$^{7}$F$^{o}$--a$^{7}$G$^{e}$ &5 	&7 	& 0.498  &  0.169     	&&5375.57 &e$^{5}$G$^{e}$--w$^{5}$G$^{o}$ & 9	  & 7	  &-1.322  & -0.904	 \\ 
	&				&	&	&	 &		 										 \\
4286.09 &z$^{7}$F$^{o}$--a$^{7}$G$^{e}$ &7 	&5 	&-0.512  & -0.874     	&&5460.80 &d$^{3}$G$^{e}$--y$^{3}$H$^{o}$ & 7	 & 9	 &-1.519  & -1.459	 \\
4286.13 &z$^{7}$F$^{o}$--a$^{7}$G$^{e}$ &7 	&7 	& 0.374  &  0.012											 \\   
4286.16 &z$^{7}$F$^{o}$--a$^{7}$G$^{e}$ &7 	&9 	& 0.705  &  0.343	&&5485.52 &d$^{3}$G$^{e}$--y$^{3}$H$^{o}$ & 9	 &11	 &-1.469  & -1.370	 \\  													  										 \\
4296.81 &z$^{7}$F$^{o}$--a$^{7}$G$^{e}$ &9 	&7 	&-0.536  & -0.921      	&&5573.42 &d$^{3}$G$^{e}$--y$^{3}$H$^{o}$ &11	 &13	 &-1.390  & -1.286	 \\
4296.85 &z$^{7}$F$^{o}$--a$^{7}$G$^{e}$ &9 	&9 	& 0.418  &  0.033      											 \\
4296.85 &z$^{7}$F$^{o}$--a$^{7}$G$^{e}$ &9 	&11 	& 0.879  &  0.494      	&&5833.94 &b$^{7}$S$^{e}$--y$^{7}$P$^{o}$ & 7	 & 9	 & 0.616  &  0.628	 \\
										 										 \\
4304.79 &z$^{7}$F$^{o}$--a$^{7}$G$^{e}$ &11	&11	& 0.377  & -0.021      	&&5999.54 &c$^{3}$I$^{e}$--t$^{3}$H$^{o}$ & 5	 & 5	 & 0.355  &  0.378	 \\
4304.75 &z$^{7}$F$^{o}$--a$^{7}$G$^{e}$ &11	&9	&-0.699  & -1.097	&&6000.09 &e$^{3}$G$^{e}$--s$^{3}$G$^{o}$ &11	 &11	 & 0.410  & -0.862	 \\
4304.77 &z$^{7}$F$^{o}$--a$^{7}$G$^{e}$ &11	&13	& 1.027  &  0.627      											 \\
	&				&	&	&	 &		&&6032.60 &b$^{5}$S$^{e}$--w$^{5}$P$^{o}$ & 5	 & 7	 & 0.497  &  0.521	 \\
4310.34 &z$^{7}$F$^{o}$--a$^{7}$G$^{e}$ &13	&11	&-1.072  & -1.482      	&&6032.67 &c$^{3}$I$^{e}$--x$^{3}$I$^{o}$ &13	 &15	 & 0.410  & -0.862	 \\
4310.36 &z$^{7}$F$^{o}$--a$^{7}$G$^{e}$ &13	&13	& 0.189  & -0.217      											 \\
4310.36 &z$^{7}$F$^{o}$--a$^{7}$G$^{e}$ &13	&15	& 1.156  &  0.747      	&&6035.52 &c$^{5}$G$^{e}$--x$^{5}$G$^{o}$ &13	  &11	  &-0.200  & -0.463	 \\
	&				&	&	&	 &		&&6036.55 &c$^{5}$G$^{e}$--x$^{5}$G$^{o}$ &13	  &13	  & 0.790  &  0.684	 \\
4352.58 &a$^{5}$P$^{e}$--z$^{5}$P$^{o}$ &5 	&3 	&-2.870  & -2.827				      							 \\	
\hline
\end{tabular}			 
}
\end{minipage}
\end{table*}

\label{lastpage}

\end{document}